\documentclass[twocolumn,appendixfloats]{emulateapj}
\usepackage{amsmath}
\usepackage{graphicx}
\usepackage{rotating}
\usepackage{longtable}

\usepackage{natbib}
\def\kms{{km s$^{-1}$}}

\def\arcmin{{\mbox{$^\prime$}}}
\def\arcsec{{\mbox{$^{\prime\prime}$}}}

\def\HI{{H\,{\sc i}}}

\shorttitle{TiNy Titans}
\shortauthors{Stierwalt et al.}

\begin{document}
\title{TiNy Titans: The Role of Dwarf-Dwarf Interactions in the Evolution of Low Mass Galaxies}
\author{S. Stierwalt\altaffilmark{1}, G. Besla\altaffilmark{2}, D. Patton\altaffilmark{3}, K. Johnson\altaffilmark{1}, N. Kallivayalil\altaffilmark{1}, M. Putman\altaffilmark{4}, G. Privon\altaffilmark{5}, G. Ross\altaffilmark{3}}

\altaffiltext{1}{Department of Astronomy, University of Virginia, P.O. Box 400325, Charlottesville, VA 22904. {\textit{e-mail:}} sabrinas@virginia.edu}
\altaffiltext{2}{Department of Astronomy, University of Arizona, 933 North Cherry Avenue, Tucson, AZ 85721}
\altaffiltext{3}{Department of Physics \& Astronomy, Trent University, 1600 West Bank Drive, Peterborough, Ontario, Canada K9J 7B8}
\altaffiltext{4}{Department of Astronomy, Columbia University, Mail Code 5246, 550 West 120th Street, New York, NY 10027}
\altaffiltext{5}{Departamento de Astronom\'{i}a Universidad de Concepci\'{o}n, Casilla 160-C, Concepci\'{o}n, Chile}

\begin{abstract}
We introduce TiNy Titans (TNT), the first systematic study of star formation and the subsequent processing of the interstellar medium in interacting dwarf galaxies. Here we present the first results from a multiwavelength observational program based on a sample of 104 dwarf galaxy pairs selected from a range of environments within the spectroscopic portion of the Sloan Digital Sky Survey and caught in various stages of interaction. The TNT dwarf pairs span mass ratios of M$_{*,1}$/M$_{*,2}$ $<$ 10, projected separations $<$ 50 kpc, and pair member masses of 7 $<$ log(M$_*$/M$_{\odot}$) $<$ 9.7. The dwarf-dwarf merger sequence, as defined by TNT at z$=$0, demonstrates conclusively and for the first time that the star formation enhancement observed for massive galaxy pairs also extends to the dwarf mass range.  Star formation is enhanced in paired dwarfs in otherwise isolated environments by factor of 2.3 ($\pm$ 0.7) at pair separations $<$ 50 kpc relative to unpaired analogs. The enhancement decreases with increasing pair separation and extends out to pair separations as large as 100 kpc. Starbursts, defined by H$\alpha$ EQW $>$ 100\AA, occur in 20\% of the TNT dwarf pairs, regardless of environment, compared to only 6-8\% of the matched unpaired dwarfs. Starbursts can be triggered throughout the merger (i.e. out to large pair separations) and not just approaching coalescence. Despite their enhanced star formation and triggered starbursts, most TNT dwarf pairs have similar gas fractions relative to unpaired dwarfs of the same stellar mass. Thus, there may be significant reservoirs of diffuse, nonstarforming neutral gas surrounding the dwarf pairs or the gas consumption timescales may be long in the starburst phase. The only TNT dwarf pairs with low gas fractions (f$_{gas} <$ 0.4) and the only dwarfs, either paired or unpaired, with H$\alpha$ EQW $<$ 2\AA\ are found near massive galaxy hosts. We conclude that dwarf-dwarf interactions are significant drivers of galaxy evolution at the low mass end, but ultimately environment is responsible for the quenching of star formation. This preliminary study is a precursor to an ongoing high resolution \HI\ and optical imaging program to constrain the spatial distribution of star formation and gas throughout the course of the dwarf-dwarf merger sequence. 
\end{abstract}

\section{Introduction}

Mergers between massive galaxies provide an important mode of galaxy evolution that can drive morphological change from gas-rich disk galaxies to red, quiescent systems, while triggering starbursts and quasar activity \citep[e.g.,][]{toomre,sandersQSO}. At low redshift, massive galaxy interactions and mergers are observed to produce diluted metals, enhanced star formation, bluer colors, and higher AGN fractions relative to their isolated counterparts \citep[e.g.,][]{armusFIR, sandersQSO, woods10, patton11, ellison11, scudder12, richmetals, patton13, ellison13}.  These observed effects are consistent with numerical simulations of merging galaxies, which predict that close pericentric passages between massive galaxies can trigger gas inflows, starbursts, and the activation of central supermassive black holes \cite[e.g.,][and references therein]{toomre, hernquist89, barnes91, mihos94, mihos96, torrey12}. 

While the processes regulating star formation and the interstellar medium (ISM) in pairs of interacting massive galaxies in the local universe have been studied extensively, similar studies have not been done for interactions between low mass, dwarf galaxies (i.e. 10$^7$M$_{\odot} <$ M$_* < 5 \times 10^9$M$_{\odot}$). Given that low mass galaxies dominate the galaxy population in the universe \citep{bing,karacat}, the majority of mergers at all redshifts are expected to be between low mass galaxies \citep[e.g.,][]{delucia, fakhouri}. Thus, the lack of observational and theoretical constraints on the nature of low mass interactions leaves a significant mode of galaxy evolution in the universe mostly uncharted.

Whether interactions between dwarf galaxies can affect their morphologies or star formation rates (SFRs) is unclear, particularly over large pair separations, given their lower masses and thus weaker tides. Dwarfs often lack stabilizing features like bars or spiral arms and are more significantly affected by supernova blowout, all of which may reduce their capacity to form stars. The forces required to funnel gas toward the center of mass of the system to trigger a central starburst may also be less effective. Moreover, low mass galaxies are often metal poor, which affects the formation of H$_2$ and thus can dramatically modify how star formation is distributed and enhanced in response to tidal torques throughout the course of a merger \citep{kuhlen}. While some massive galaxy mergers may end in quenched systems \citep[e.g.][]{cox06, galzooquenched}, $<$2\% of isolated galaxies with stellar masses $<$10$^9$M$_{\odot}$ are non-star forming \citep{geha12}. Thus galaxy mergers could proceed differently at lower masses.

Dwarf galaxies are known to undergo starbursts, although the origins of these starbursts are unknown \citep{heckmanSB, johnsonHaro,SBmcquinnI}. Furthermore, studies of resolved stellar populations in nearby dwarfs find their star formation histories (SFHs) to be inconsistent with simple closed box models \citep[i.e. single bursts, constant or smooth, exponentially declining SFRs;][]{weiszANGST}. Mergers between dwarf galaxies have been proposed as the explanation for the blue compact dwarf (BCD) class of galaxy, which are characterized by intense starburst activity \citep{bekki,lelliSB1,lelliSB2,kolevaSB}. However, a systematic study of the starburst properties of paired dwarfs versus their isolated counterparts is critical for disentangling merger-driven starbursts and those due to stochastic processes.

Individual examples of local dwarf pairs (D $<$ 20 Mpc) provide further support for the importance of dwarf-dwarf interactions in the evolution of low mass galaxies. A classic example is the Magellanic-type galaxy pair NGC 4485/4490 which is surrounded by a large HI envelope with a gaseous bridge that connects the two dwarfs \citep{clemens98}. Despite the fact that the dwarfs have not yet coalesced (they are separated by a projected distance of 10 kpc), they show disturbed morphologies and NGC4485 is undergoing a starburst, plausibly triggered by the interaction \citep{11HUGS}. This system bears striking resemblance to our closest example of an interacting pair of dwarfs, the Magellanic Clouds.  Connected by a gaseous bridge and trailed by a massive HI complex known as the Magellanic Stream, the Clouds illustrate clearly that interactions between dwarf galaxies can play an important role in the redistribution of both gaseous and stellar material \citep{besla12}. 

To probe this so far mostly unconstrained mode of galaxy evolution and in light of the ubiquity of dwarf-dwarf mergers across cosmic time, we are conducting the first systematic observational and theoretical study of star formation and the subsequent processing of the ISM in interacting dwarfs as a population: TiNy Titans (TNT). This paper describes the observational portion of TNT, a sample of 60 isolated dwarf galaxy pairs selected from the spectroscopic portion of Data Release 7 of the Sloan Digital Sky Survey (SDSS) and caught in various stages of interaction. Each dwarf pair member has a stellar mass within 10$^7$M$_{\odot} <$ M$_* < 5 \times$10$^9$M$_{\odot}$ and is within a redshift range of 0.005 $<$ z $<$ 0.07. 

To disentangle the effects of the dwarf-dwarf interaction from any large scale environmental effects, all 60 TNT dwarf pairs are also selected to be isolated, or more than 1.5 Mpc from a more massive galaxy (M$_* > 5 \times$10$^9$M$_{\odot}$). \cite{geha12} found that the fraction of quenched low mass galaxies was negligible at such separations, providing a clean spatial delineation for when environmental factors dominate their evolution. To further control for the effects of massive perturbers that may enhance or dilute trends observed for the isolated pairs, TNT also includes 44 nonisolated pairs that cover the same range of stellar masses and redshifts, but are found within 1.5 Mpc of a massive host.  

In Section \ref{goals}, we outline the main goals of TNT, the first of which is to define the dwarf-dwarf merger sequence at z$=$0. In Section \ref{select}, we describe in detail the selection of our sample from the SDSS and place our pairs in the context of two well-studied dwarf interactions, NGC4485/4490 and the Large and Small Magellanic Clouds (LMC/SMC). We also describe in detail the comparison samples we use in our analysis (i.e. nonisolated pairs and isolated single dwarfs) to control for environmental effects and stochastic processes. We discuss our re-reduction of optical data from the SDSS as well as new observations (and related data reduction) of the \HI\ content of our dwarf pairs in Section \ref{obs}. In Section \ref{sfr}, we present the main results from the SDSS imaging and spectroscopy, specifically the effects of dwarf-dwarf interactions on SFR and the frequency at which they trigger starbursts for both our isolated and nonisolated pairs. In Section \ref{gasfrac}, we explore the evolution of the gas fractions observed for dwarfs throughout the merger sequence and as it relates to their environment. Finally, we discuss the implications of our results in Section \ref{disc}.

\section{Scientific Objectives of TiNy Titans}\label{goals}
There are a rapidly growing number of surveys of dwarf galaxies using NASA observatories, including the ANGST HST survey \citep{dalcantonANGST,weiszANGST}, the Herschel Dwarf Galaxy Survey \citep{maddenHDGS}, the COS-Dwarfs Survey \citep{bordoloiCOS}, as well as the Local Volume Legacy Survey and 11-Mpc H$\alpha$ Ultraviolet Galaxy Survey \citep[LVL \& 11HUGS;][]{kennicuttLVL,LeeLVL}. Despite these numerous panchromatic surveys, none are focused on the impact of interactions between dwarf galaxies on their evolution. TNT will address this void by combining both a multiwavelength observational effort and a theoretical approach using large-volume cosmological simulations and hydrodynamic simulations of individual systems to provide statistics not possible with only a small number of known dwarf galaxy pairs. Specifically, we aim 1) to establish the dwarf-dwarf merger sequence at z$=$0; 2) to connect interacting dwarfs at the current epoch with high redshift analogs; and 3) to probe interaction-driven modes of star formation.\\

\noindent {\bf{Defining the dwarf-dwarf merger sequence:}} 
First, we aim to define the dwarf-dwarf merger sequence at z $=$ 0 using the metrics of star formation rates, gas fractions and signs of tidal distortion. Using a careful selection process (described in Section \ref{select}), we define a sample of 60 isolated dwarf pairs that are caught in various stages of interaction. This large and systematically selected sample will allow us to study dwarf-dwarf interactions and mergers as a population and to determine how much of what we know about mergers between massive galaxies can be scaled down to lower masses. A more detailed look at morphology and tidal distortions based on high resolution \HI\ imaging and optical asymmetries in the SDSS imaging as a function of pair separation will be examined in an upcoming paper (Stierwalt et al. $in~prep$). \\

\noindent {\bf{Nearby dwarf pairs as windows to high redshift:}}
Second, the interaction-triggered star formation occuring in our sample of low metallicity dwarf pairs provides a unique window into the hierarchical processes that should operate more frequently at higher redshifts. For example, high resolution studies at high redshift (z$\sim$1-2) find star formation to be typically off-center and significantly clumpier than normal, star-forming galaxies at the current epoch \citep{wuyts12,wuyts13}. However, the low resolution SDSS optical imaging available for the TNT dwarf pairs already suggest their star formation is qualitatively similar to that of the clumpy, irregular massive galaxies at high redshift. Until we can resolve dwarf galaxy interactions at high z, their low redshift analogs (i.e. low metallicity, dwarf pairs) offer the best insight into the modes of star formation at earlier epochs.\\

\noindent {\bf{Interaction-driven modes of star formation:}}
Third, the TNT survey will explore extreme modes of star formation triggered by tidal interactions at low metallicity, including star formation along tidal bridges and tails. By including a large number of dwarf pairs involved in a variety of interaction stages, we will constrain the frequency of analogs to the large HII region, 30 Doradus, in the LMC and at what stages of the interaction they may occur. As 30 Doradus is the most extreme HII region in the Local Group, the prevalence of such structures is crucial to understanding the energetics and scale of star forming regions in dwarf galaxies, and by analogy, in galaxies at higher redshifts. To enable this effort, we have an ongoing program to obtain deep broad- and narrowband (H$\alpha$) optical imaging to resolve the distribution of star formation across each TNT dwarf pair and along any tidal structures that exist (Kallivayalil et al. $in~prep$).\\

Finally, we will address many of these goals from a theoretical standpoint.
Only now are simulations of large cosmological volumes able to probe mass scales relevant to the evolution of dwarf galaxies (e.g. mass scales of the Large Magellanic Cloud down to the Sagittarius Dwarf). In a complementary study, we will utilize state-of-the art cosmological simulations to quanitfy the frequency and orbital properties of dwarf groups in concordance $\Lambda$ Cold Dark Matter cosmology ($\Lambda$-CDM; Besla et al. in prep).  By comparing such statistics to the properties of dwarf groups observed in surveys such as SDSS, our theoretical program will introduce new tests for $\Lambda$-CDM predictions at the low mass scale.

\section{A Systematic Study of Dwarf Pairs}\label{select}
In order to probe the star formation and ISM physics driven by  dwarf-dwarf interactions and the frequency of these encounters in the field, we first identify a sample of potentially interacting dwarfs that is complete within the limitations of the SDSS. In the remainder of this section we describe this selection in detail and compare the resulting sample with two well-studied interacting dwarf pairs, NGC4485/4490 and the Magellanic Clouds.

\subsection{TiNy Titans Isolated Dwarf Pairs}We use the spectroscopic portion of SDSS Data Release 7 to identify nearby (0.005 $<$ z $<$ 0.07), low-mass (10$^7$M$_{\odot} <$ M$_* < 5\times10^9$M$_{\odot}$) galaxies determined to be in pairs based on their small projected radial separation (R$_{sep} <$ 50 kpc) and small line of sight velocity separation (v$_{sep}$ $<$ 300 \kms). 

\begin{figure}[h!]
\begin{center}
\includegraphics[height=2.4in,width=3.5in]{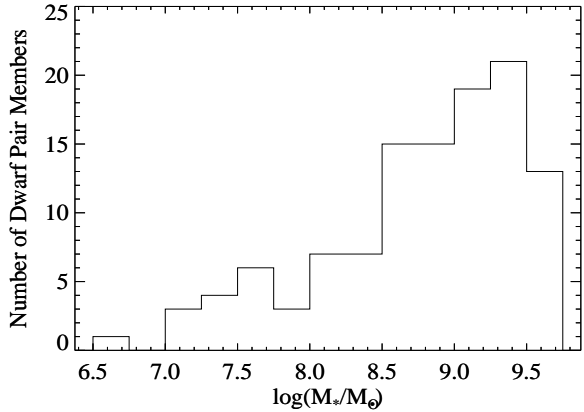}
\caption{Distribution of the logarithm of the stellar mass for all 120 isolated dwarf pair members. Stellar masses are determined from re-measured broadband ugriz photometry from SDSS images as described in Section \ref{sdssreduct}.
\label{mstarhisto}}
\end{center}
\end{figure}

These limits in R$_{sep}$ and v$_{sep}$ are similar to those known to identify interactions in SDSS-selected samples of more massive galaxies \citep[e.g.,][]{ellison11, scudder12, patton13}. For example, \cite{patton13} find star formation to be enhanced in pairs of massive galaxies out to separations of $\sim$100 kpc, which is approximately half the virial radius for a galaxy with a dark matter halo mass of $\sim$10$^{12}$M$_{\odot}$. The TNT pair members have an average stellar mass of M$_* \sim $10$^9$M$_{\odot}$ (see Figure \ref{mstarhisto}), which translates to a dark matter halo mass of $\sim$10$^{11}$M$_{\odot}$ at z$=$0 \citep{moster}. The corresponding virial radius is $\sim$120 kpc (assuming a virial overdensity $\Delta_{vir} = $360 and $\Omega_M =$ 0.27). The upper limit placed on R$_{sep}$ for our selection is thus approximately half the virial radius and analogous to the spatial scales of relevance for massive galaxy pairs. Despite the large upper limit on the velocity separation, only six of our 60 isolated pairs have v$_{sep} > $150 \kms, indicating that the majority of the TNT pairs are likely bound systems, although further information on transverse velocities is needed. Finally, the selected ranges of R$_{sep}$ and v$_{sep}$ are further justified by the low resolution SDSS imaging which reveal that some dwarf pairs at separations of 50kpc still show faint stellar tidal features suggestive of interactions. 

Our pair-finding algorithm follows the method outlined in \cite{patton13}. For each galaxy that satisfies the stellar mass, redshift, and isolation criteria, a search is done for a companion that satisfies the radial and velocity separation criteria. The closest companion is selected and then both the primary and the secondary pair members are removed from future search efforts to avoid double counting. Thus, some of our pairs may, in fact, be part of triple systems. The dwarf group environment and the frequency of such multiple systems will be discussed in a companion paper (Besla et al. $in~prep$). Instead we focus here on the closest companion to each dwarf.

To confine our study to interactions that can plausibly affect the structure and SFR of both the primary and the secondary pair members, we further restrict our dwarf pairs to have stellar mass ratios of less than 10 (i.e. (M$_1$/M$_2$)$_* < $10 where M$_1$ is the mass of the primary). This limit on mass ratio also allows for a more complete sample since the sensitivity of SDSS to a much lower mass secondary pair member falls off quickly as a function of redshift. Despite this range of allowed mass ratios, only four of our 60 pairs have (M$_1$/M$_2$)$_* > $5. Due to our restriction on stellar mass and stellar mass ratio, our sample does not include interacting systems like the starburst NGC4449 and its very faint, low surface brightness companion \citep[(M$_1$/M$_2$) $>$ 50;][]{md4449}. Instead, as shown in Figure \ref{mstarhisto}, the median stellar mass for our pair members is log(M$_*$/M$_{\odot}$) $=$ 8.9.

To disentangle the effects of dwarf-dwarf interactions from those resulting from other environmental factors, like ram pressure stripping or tidal forces from a more massive galaxy, we select only dwarf pairs that are farther than 1.5 Mpc from a more massive neighbor (defined as having M$_* > 5 \times 10^9$M$_{\odot}$). Massive galaxy neighbors are observed to significantly influence dwarf galaxy evolution in the form of the well known distance-morphology relationship wherein gas-poor, quenched dwarf spheroidals (dSphs) are preferentially located closer to massive galaxy group members \citep[e.g.,][]{morphseg, stierwalt09, morphsegWeisz}. Our isolation criterion is even stricter than that set by \cite{geha12} who found that all dwarfs farther than 1.5 Mpc from an L$^*$ host (i.e. M$_* > 2.5 \times 10^{10}$M$_{\odot}$) are readily forming stars and show no signs of quenching. Our pairs are selected from the larger sample of galaxy pairs of \cite{patton13}, and thus the stellar masses used in our initial sample selection come from \cite{mendel} and rely in part on the photometry of \cite{simard}. To ensure our pairs are truly isolated, we also use the NASA-Sloan Atlas\footnote{http://www.nsatlas.org/} to search for massive neighbors and more accurately determine environment. 

Five of our pairs are within 1.5 Mpc of the SDSS boundary. We verify these pairs are isolated by searching for possible massive neighbors in the 2MASS Extended Source Catalog. We determine redshifts for any galaxies within 1.5 Mpc but outside of the SDSS footprint and with M$_{K_s} < - 21$ (which translates to M$_* > 5\times10^9$M$_{\odot}$ assuming (M/L)$_{K_s} = 1$M$_{\odot}$/L$_{\odot}$) using the NASA Extragalactic Database (NED) and the 40\% ALFALFA catalog \citep[$\alpha$40;][]{a40}. All five pairs are found to still satisfy our isolation criteria.

\begin{figure}[h!]
\begin{center}
\includegraphics[height=2.4in,width=3.5in, viewport=20 0 500 345,clip]{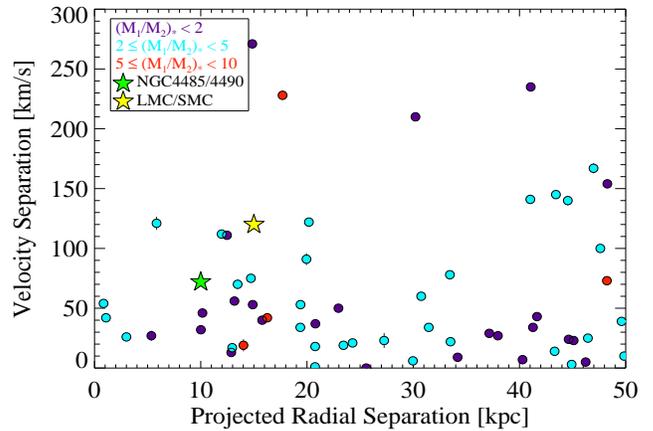}
\caption{TiNy Titans Sample Selection: Projected radial and line of sight velocity separations for the 60 TNT isolated
  dwarf pairs color-coded by stellar mass ratio between
  the two dwarfs. Although we allow pairs to have stellar mass ratios up
  to 10 and velocity separations up to 300 \kms, only four pairs have (M$_1$/M$_2$)$_* > $5 (red points) and only six pairs have v$_{sep} > $ 150 \kms. The radial and velocity separations of the interacting dwarf pairs, NGC4485/4490 (green star) and the Large and Small Magellanic Clouds (yellow star), which both have stellar mass ratios of M$_1$/M$_2$)$_* \sim$9, are marked for reference only, as they are not included in the TNT sample.
\label{selectfig}}
\end{center}
\end{figure}

\begin{figure*}
\begin{center}
\includegraphics[height=2.4in,width=2.4in]{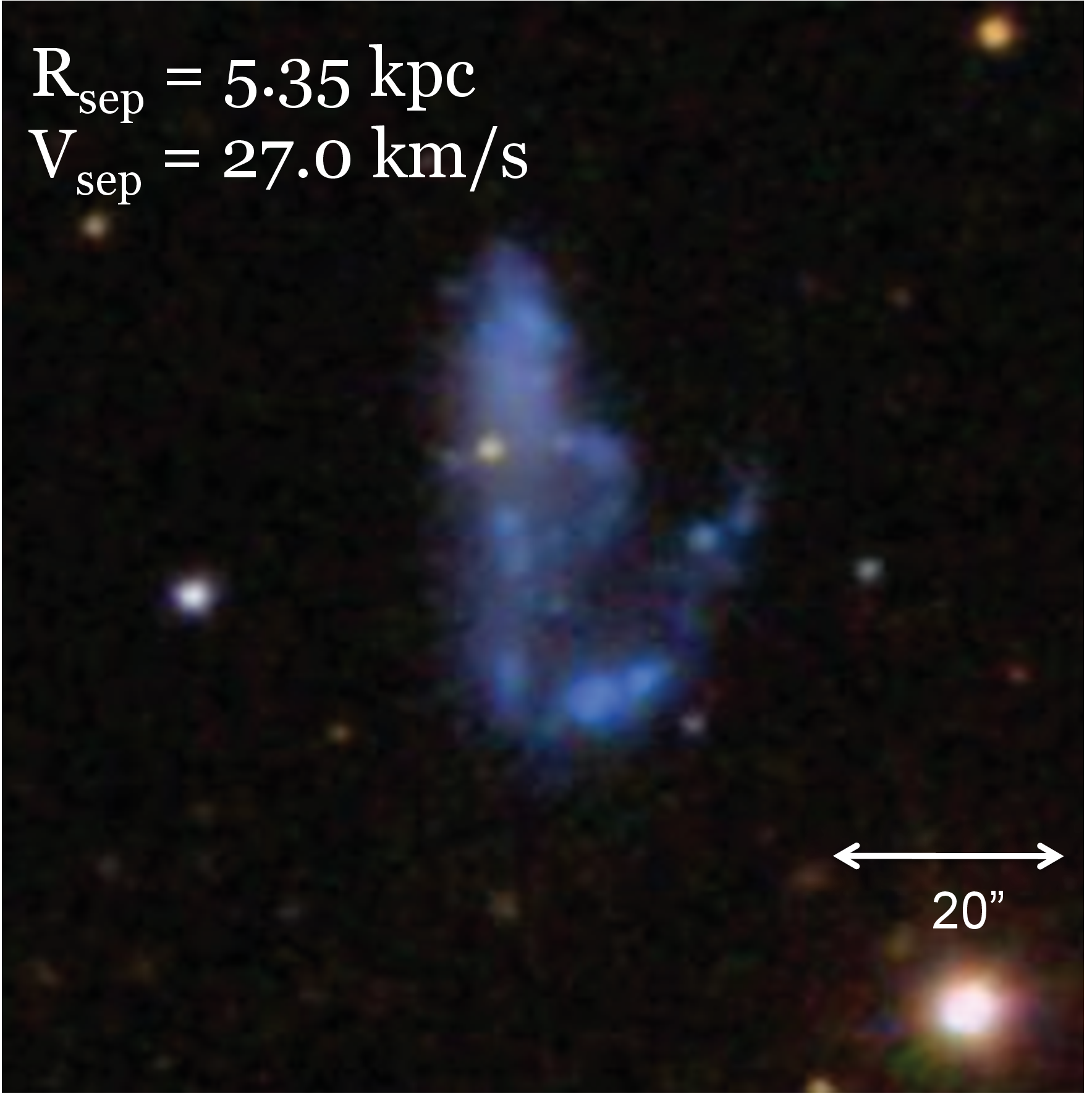}
\includegraphics[height=2.4in,width=2.4in]{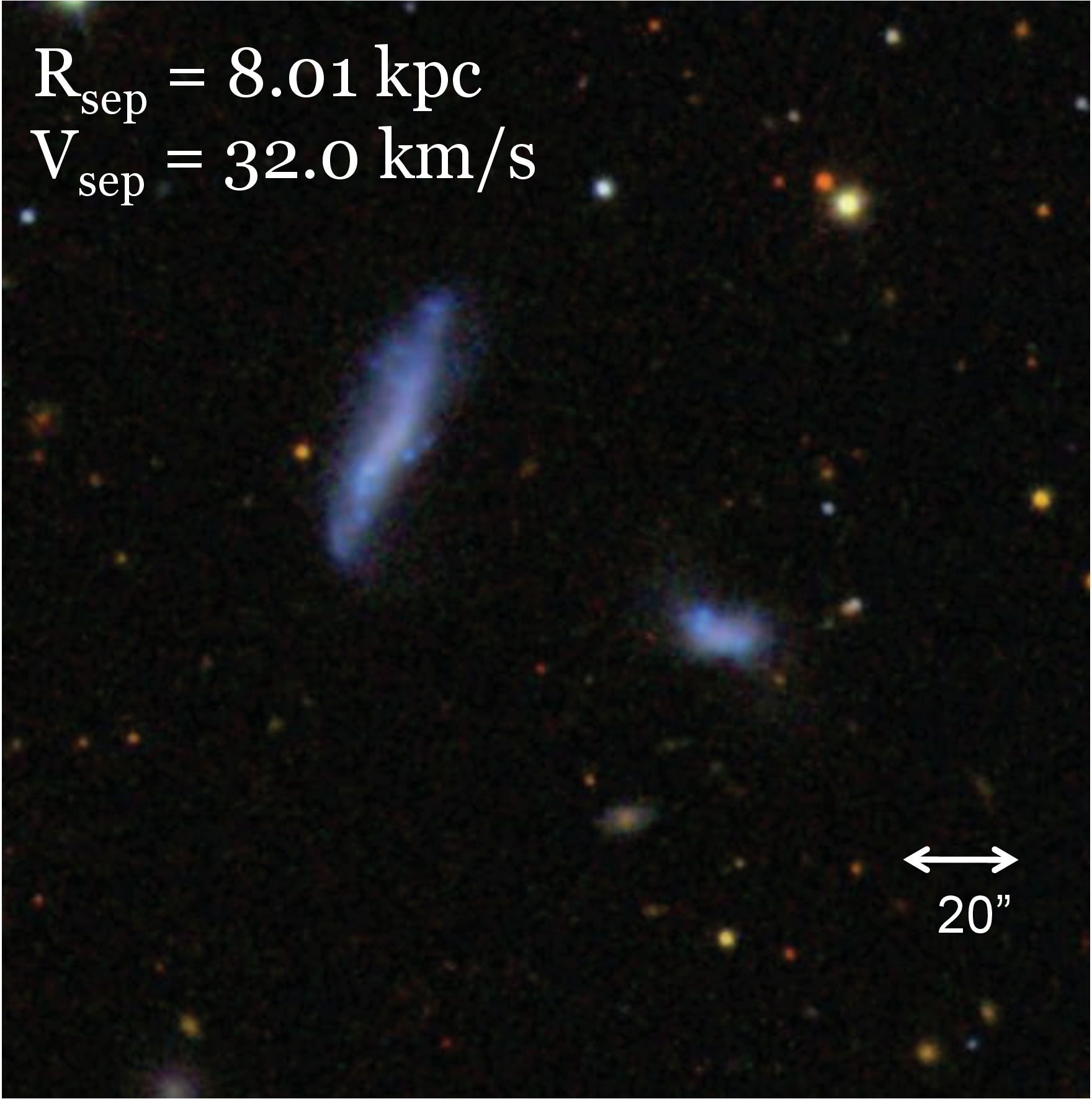}
\includegraphics[height=2.4in,width=2.4in]{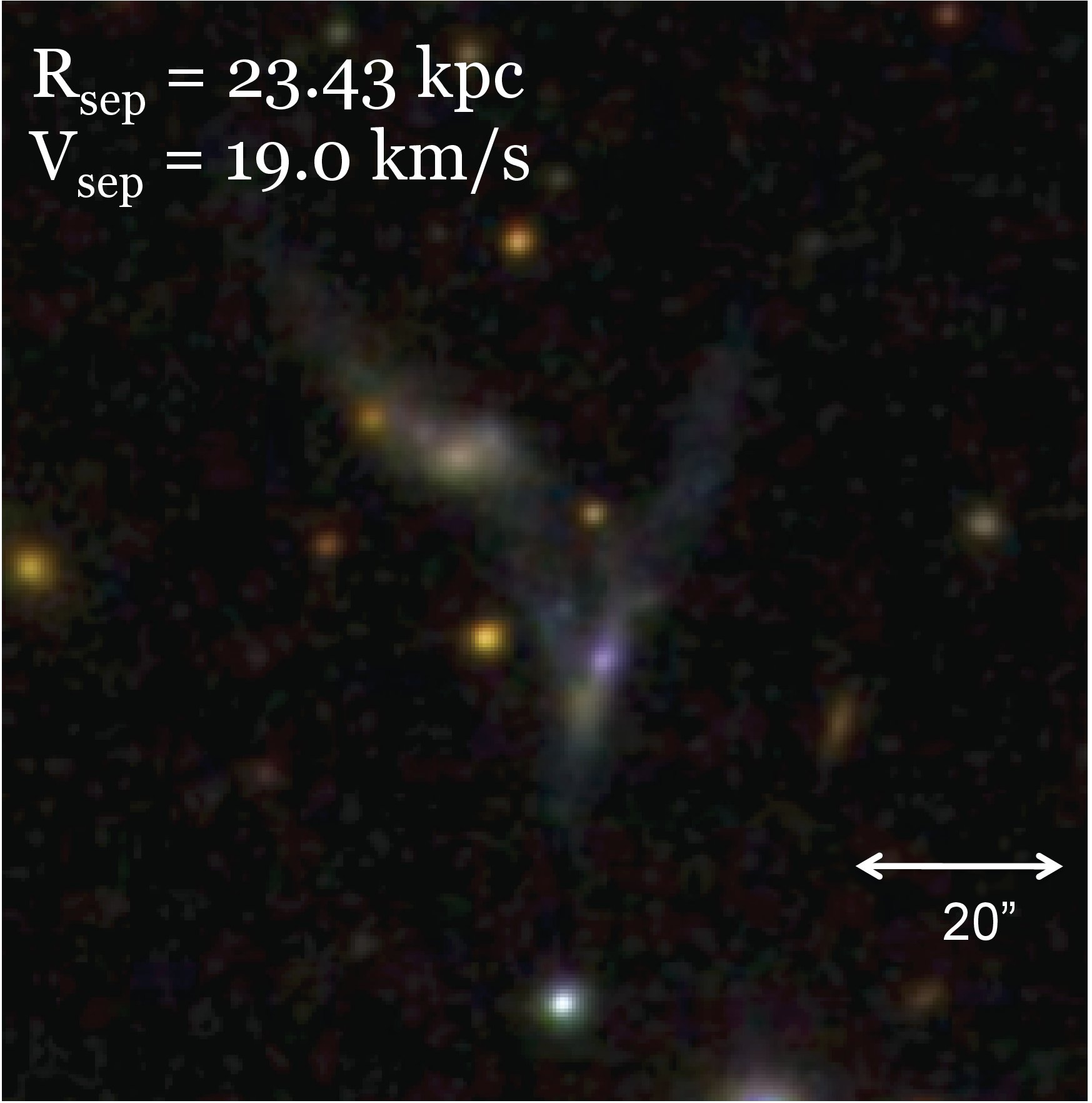}
\includegraphics[height=2.4in,width=2.4in]{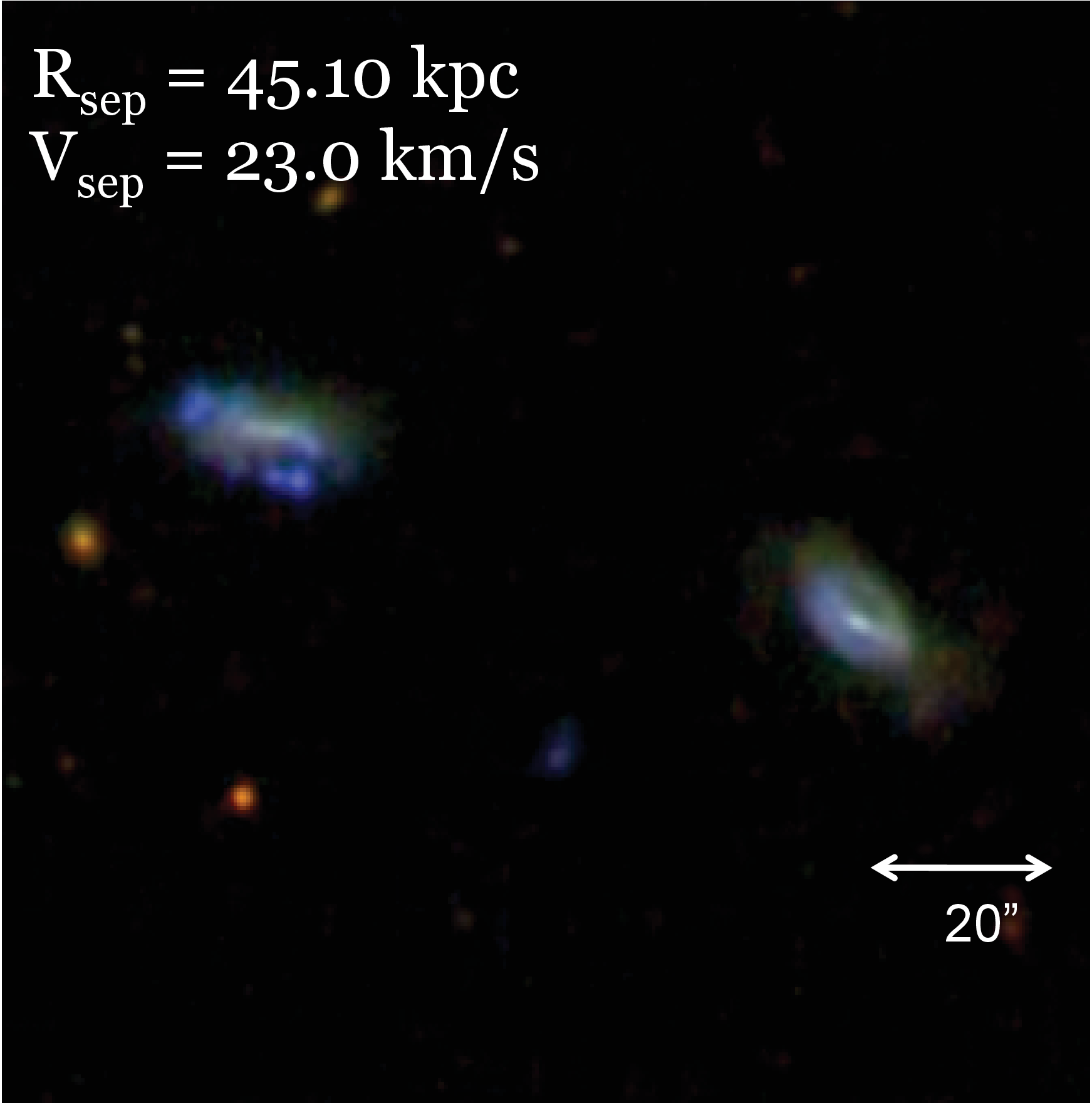}
\caption{Broadband ugriz SDSS images of four isolated dwarf pairs from TiNy Titans covering a range of projected radial separations (R$_{sep}$), which are noted along with the velocity separation (v$_{sep}$) for each pair. The disturbed morphologies reveal stellar bridges and tails, as well as asymmetrically-distributed blue star forming knots. Given the isolation of these pairs ($>$ 1.5 Mpc from a more massive galaxy), these distortions are not the result of large scale environmental effects and are likely due to the dwarf-dwarf interaction. Optical images and \HI\ emission spectra for all 60 TNT isolated pairs are shown in Figure \ref{atlas}.
\label{examples}}
\end{center}
\end{figure*}

Our selection criteria result in a sample of 60 dwarf pairs probing a range of radial separations and thus sampling what we have defined as the dwarf-dwarf merger sequence. The distribution of the TNT isolated pairs in terms of our three main selection criteria (after stellar mass), r$_{sep}$, v$_{sep}$, and stellar mass ratio, is shown in Figure \ref{selectfig}. The majority of our pairs show evidence of an interaction as indicated by disturbed or irregular optical morphologies in the SDSS imaging. Examples are highlighted in Figure \ref{examples}.

By selecting our sample from the spectroscopic portion of SDSS, we greatly reduce contamination from projected (but not actual) pairs with the additional redshift information. However, this selection limits our study to dwarfs with Petrosian $r$-band magnitudes of r $<$17.77. The overall completeness of the spectroscopic coverage of the main galaxy sample is estimated to be $\sim$88\% \citep{PA08}. Additionally, all pairs with separations of less than 55\arcsec~would be missed due to spectroscopic fiber collisions \citep{sdssspect}. Fortunately, due to overlapping and multiple tiles of spectroscopic coverage, $\sim1/3$ of these close pairs are recovered. \citep[See Figure 2 of][for a comparison of the number of spectroscopic pairs to photometric pairs that do not suffer the same fiber placement issue.]{PA08} In fact, half of the TNT pairs have $\theta_{sep} <$ 55\arcsec, and, as a result, our sample includes five pairs with r$_{sep} < $10 kpc. However, our selection criteria are still biased toward more clearly separated pairs, or merger stages prior to a final coalescence. The closest pair has a projected separation of 0.85 kpc at a distance of 31 Mpc which translates to a separation of 5.65\arcsec\ on the sky and approaches the minimum separation at which the two nuclei are still discernible.

The median velocity uncertainty of the SDSS spectra for all of our pairs is $\pm$37 \kms, and most (89\%) of the TNT pairs have velocity uncertainties of $<$100 \kms. Thus, we do not expect significant contamination from false pairs due to large velocity uncertainties, especially given that most of our pairs have v$_{sep} <$ 150 \kms.

\subsection{Two Well-Known Interacting Dwarf Pairs}\label{knownpairs}

The Magellanic pair NGC4485/4490, a well-studied pair of interacting dwarfs, is connected by an HI, star forming bridge \citep{smith4490}, analogous to the HI in the Magellanic Bridge \citep{kerr, putmanstream}. Due to the relative isolation of the pair from more massive galaxies \citep{karacat}, the disturbed optical and \HI\ morphologies are likely a result of the interaction between the pair. The interaction appears to have triggered a starburst (EQW(H$\alpha$+[NII]) = 76$\AA$; SFR = 0.14 M$_{\odot}$ yr$^{-1}$; \cite{11HUGS}) in the smaller galaxy NGC4485. Knots of star formation are observed on the side nearest the more massive dwarf NGC4490, which is in turn substantially warped. The pair is embedded in one of the largest \HI\ envelopes known \citep{clemens98} with a total \HI\ mass of M$_{HI} =$1.8$\times$10$^{10}$M$_{\odot}$ \citep{hucht4490}. Using the absolute B band magnitudes from \cite{kennicuttLVL} (M$_B =-$16.97 for NGC4485 and M$_B =-$19.35 for NGC4490), a distance of 7.8 Mpc \citep{tully88}, and assuming a mass-to-light ratio of (M/L)$_B =$1.03 \citep{bell03}, the resulting stellar masses (9.8$\times$10$^8$ M$_{\odot}$ and 8.6$\times$10$^9$ M$_{\odot}$ for NGC4485 and NGC4490 respectively) place NGC4490 slightly above the cut off for the TNT sample selection. The resulting gas fraction of f$_{gas} = $1.4M$_{HI}$/(1.4M$_{HI}$ + M$_*$) $=$0.72 for the pair is unusually high for galaxies in this stellar mass range \citep{huang12}. 

Another pair of well-studied interacting dwarf galaxies is the Large and Small Magellanic Clouds (LMC/SMC), but any effects from an LMC-SMC interaction on their star formation or gas content are difficult to disentangle from effects resulting from their proximity to the Milky Way. At a projected separation of 15 kpc, a tidal interaction between the LMC and SMC may explain why the LMC is bluer on average than analogs in the field \citep{tollerudLMC}. If the Clouds have been accreted fairly recently \citep{kalli13}, their large connecting bridge of \HI~gas \citep[Magellanic Bridge;][]{kerr} is most likely the result of a tidal interaction between the two dwarfs themselves, rather than with the Milky Way \citep{besla10, besla12}. The massive trailing \HI~stream associated with the pair \citep[Magellanic Stream;][]{putmanstream} is likely extended along the orbit of the Clouds by long range tides from the Milky Way, but may have been initially seeded from an LMC-SMC interaction. Global tides from a massive host can also induce star formation in a satellite dwarf galaxy at pericentric passages. Thus, disentangling the influence of a low mass companion from that of a massive host resulting in off-centered star formation can be difficult. Neither the SMC or LMC are classified as starbursts with SFR $=$ 0.05 M$_{\odot}$ yr$^{-1}$ \citep{SMCSFR} and SFR $\sim$ 0.2 M$_{\odot}$ yr$^{-1}$ \citep{LMCSFR}, respectively. Assuming M$_* = $3.1$\times$10$^8$M$_{\odot}$ for the SMC \citep{stanSMC}, M$_* = $2.7$\times$10$^9$M$_{\odot}$ for the LMC \citep{vdmLMC}, and M$_{HI} = $9$\times$10$^8$M$_{\odot}$ for the pair (including the Bridge) \citep{putmanstream}, $f_{gas}$ for the LMC/SMC system is only 0.3, much lower than the gas fraction observed for the more isolated NGC4485/4490. Some of the ``missing'', potentially stripped, gas may still be present in much more diffuse features like the Magellanic Stream, and, as shown by \cite{fox14}, much of the gas has been ionized due to collisions with ambient gas in the Milky Way's gaseous halo.

Despite the detailed studies of these few intriguing examples, very little is known about whether these systems are representative of dwarf-dwarf interactions in general or how the observed effects on their star formation histories and ISM vary with interaction stage. With the TNT sample we seek to explore the following questions: Over what range of pair separations is star formation enhanced in each pair member? How common is the starburst phase for interacting dwarfs? Why is a burst induced in NGC4485 but not in the SMC? What are the \HI\ properties of dwarf pairs? Are most dwarf pairs surrounded by large gaseous envelopes like both NGC4485/4490 and the Clouds? Over what range of pair separations are stellar and/or \HI\ bridges and warps observed? 

At projected separations of R$_{sep} = $10 kpc (15 kpc) and relative line of sight velocities v$_{sep} = 72$ \kms\ (120 \kms) for NGC4485/4490 (LMC/SMC), these two pairs are comparable to pairs with the smallest projected separations in the TNT sample. However, given their high stellar mass ratios of (M$_1$/M$_2$)$_* \sim$9, similar pairs are less likely (but not impossible) to be found in the TNT sample owing to the sharply decreasing SDSS sensitivity with redshift, which makes the smaller companion increasingly difficult to detect at greater distances. The prevalence of such high mass ratio pairs will be discussed in more detail in comparison to cosmological expectations in the companion paper (Besla et al. $in~prep$). 

\subsection{Control Sample Selection}\label{controls}
In order to isolate the role of dwarf-dwarf interactions in the evolution of low mass galaxies, we define and make use of multiple comparison samples to control for potentially competing effects due to 1) nearby massive perturbers, 2) stochastic internal processes, and 3) changes in pair separation. We describe each of these control samples in detail in the remainder of this section. \\

\noindent {\bf{TNT Nonisolated Pairs:}} To establish trends among dwarf-dwarf interactions that are independent of large scale environment, TNT also includes a representative sample of 44 dwarf pairs near (within 1.5 Mpc) a massive neighbor (M$_* > $5$\times$10$^9$M$_{\odot}$). These nonisolated pairs otherwise follow the same pair selection criteria as our isolated sample, including stellar mass, redshift, mass ratio, and radial and velocity separation limits. We have obtained new single-dish \HI\ observations and re-measured optical photometry to derive stellar and \HI\ masses for these pairs using the same techniques described in Sections \ref{sdssreduct} \& \ref{HIreduct}. We refer to this sample throughout the remainder of the paper as the {\it{TNT nonisolated pairs}}.\\

\noindent {\bf{Matched Isolated/Nonisolated Single Dwarfs:}} To highlight any changes in gas fraction or SFR augmentation resulting from a dwarf-dwarf interaction, we also compare both our isolated and nonisolated interacting dwarf pairs to $matched$ sets of isolated and nonisolated unpaired dwarfs also selected from the spectroscopic portion of the SDSS. For each dwarf pair member, both from the isolated and nonisolated pairs, we search for unpaired dwarfs of similar stellar mass, redshift, local density, and isolation following the control sample methodology of \cite{patton13}. All unpaired dwarfs that provide a good match are kept and used in our analysis with the additional constraint that each dwarf must be $>$150 kpc from another dwarf (i.e. M$_* < $5$\times$10$^9$M$_{\odot}$) in an attempt to remove dwarf pairs and groups. Although we assume the resulting sample of 480 isolated and 664 nonisolated dwarfs to be unpaired, low mass companions that fall below the sensitivity of the SDSS observations (a strong function of redshift) could still be present. We refer to this sample throughout the remainder of the paper as the {\it{matched isolated/nonisolated single dwarfs}}. 

To constrain the neutral gas content of our matched single dwarf sample, we search for each SDSS-selected dwarf in the Arecibo-based ALFALFA $\alpha$40 catalog \citep{a40}. Of the 480 (664) isolated (nonisolated) matched single dwarfs, 214 (290) fall within the sky coverage of the ALFALFA $\alpha$40 catalog, and 97 (138) are detected in \HI. The majority ($\sim$80\%) of the control galaxies that fall within the ALFALFA coverage, but are not detected, have z$>$0.025, where the ALFALFA detection limit rises to M$_{HI}\sim$10$^9$ M$_{\odot}$ \citep{a40}. Among the control galaxies, the \HI\ nondetections follow the same distributions in H$\alpha$ equivalent width, SFR, and stellar mass as those control galaxies that are detected in \HI, suggesting that any control comparison involving \HI\ mass is likely only biased toward more nearby galaxies or towards more HI-massive galaxies at z$>$0.025.\\ 

\noindent {\bf{Extended Pair Separation Samples:}} When determining changes in SFR as a function of pair separation (see Section \ref{sfr}), we extend both our isolated and nonisolated pair samples to include pairs with separations out to 300 kpc while keeping all other selection criteria the same. These larger samples allow us to verify the extent to which any observed effects occur beyond the 50 kpc separations of the main sample. Thus, the paired dwarfs at the lowest radial separations in the extended pair sample (i.e. R$_{sep} < $50 kpc), are identical to the 60 TNT isolated pairs with a few minor differences. For example, pairs within 1.5 Mpc of the boundary of the SDSS footprint are not included in the extended pair sample, whereas they are included (with additional scrutiny, as described in Section \ref{select}) in the main TNT sample. Additionally, only pairs with reliable photometry in \cite{simard}, and thus stellar masses, for both pair members were considered for the main TNT sample, but we include in Figure \ref{sfrenhance} individual pair members even when photometry for the other pair member is unreliable. (See Figure 11 of \cite{simard} and the associated text for a description of how the quality of the photometry was quantified.)

\section{Observations \& Data Reduction for TiNy Titans}\label{obs}

Here we present the first observational results for TNT which draw upon global photometry and star formation rates from SDSS imaging and spectroscopy, as well as global neutral gas content from new single-dish \HI~observations using the Green Bank and Arecibo Telescopes.

\subsection{Broadband Optical Colors, Stellar Masses,  \& SFRs from SDSS Photometry \& Spectroscopy}\label{sdssreduct}
In order to properly assess galaxy environment, we require reliable stellar masses for a large majority of SDSS galaxies to use in our sample selection. Thus, for the initial selection of the TNT sample (and the determination of the stellar mass ratios in Figure \ref{selectfig}), we use the stellar masses of \cite{mendel} which are derived by applying a dusty model to the photometry of \cite{simard}. However, foreground stars and background galaxies can sometimes contaminate the photometry of nearby galaxies, especially dwarfs, and these special cases are not always caught by the large \cite{simard} and \cite{mendel} catalogs. Thus, for the remainder of our analysis, we perform our own photometry using downloaded, calibrated SDSS images and derive stellar masses for the 104 TNT isolated and nonisolated pairs from the resulting colors. 

To determine reliable stellar masses for our sample, we begin by measuring the brightness of each dwarf by summing the flux within a contour tracing the 2-$\sigma$ noise level (usually 0.03 to 0.04 nanomaggies) as determined by the r-band image. We apply the same aperture to the remaining four bands and use the Galactic dust maps of \cite{dirbe} from the COBE/DIRBE satellite to determine the galactic extinction correction to be applied along each line of sight and for each SDSS filter. We then derive stellar masses by taking advantage of the SED fitting built into the {\tt{kcorrect}} software \citep{kcorrect} but without applying a k-correction to our (low redshift) dwarfs. The software assumes a \cite{chabrierIMF} initial mass function. Our resulting stellar masses agree within 10\% for most (87\%) of the galaxy pair members but differ by up to 60\% in cases where we apply a truncated aperture in order to avoid a foreground or background object. 

SFRs for SDSS-selected galaxies are pulled from \cite{jarle04}. Similar to the strategy applied by \cite{scudder12} and \cite{patton13}, we only include in our SFR analysis those galaxies classified as star forming based on the criteria of \cite{kauff} and require a signal-to-noise $>$1 in each emission line used in this classification. The SFRs are derived from the emission within the spectrum and thus depend on the placement and extent of the SDSS fiber. For the majority of pair members, the fiber falls on the peak and near the center of the light distribution and thus provides a look at the brightest knot or portion of each dwarf. The covering fraction for the TNT pair members (i.e. the ratio of the flux observed within the fiber to the total flux) has a median of 17\% and ranges from $\sim$2-55\%. 

\subsection{Global HI Content from GBT \& Arecibo}\label{HIreduct}
We obtained global \HI\ mass measurements for 89 and \HI\ mass upper limits for 11 of our 60 isolated and 44 nonisolated dwarf pairs. New observations for 87 of these pairs were made with either the 100-m Green Bank Telescope (GBT) or the Arecibo 305-m single dish radio telescope over multiple observing runs from February 2013 to May 2014. The remaining 13 pairs have publicly available \HI\ line spectra in the Cornell Digital \HI\ Archive\footnote{http://arecibo.tc.cornell.edu/hiarchive/}. Due to the large beam sizes ($\sim$9\arcmin~for GBT and 3\arcmin.5 for Arecibo), the resulting global \HI\ spectra are sensitive to the total \HI\ content of each pair system, including diffuse structures that may be in extended gaseous envelopes like those surrounding NGC4485/90 and the Magellanic Clouds \citep{clemens98, putmanstream, kerr}. Despite the large beam sizes, our isolation criteria guarantees that no galaxies with M$_* >$5$\times$10$^9$M$_{\odot}$ fall within either single dish beam. Even for the most distant pair in our sample (300 Mpc), the 9\arcmin~beam translates to $\sim$0.8 Mpc which is less than the isolation requirement of 1.5 Mpc. While there is a possibility the beam may pick up emission from other neighboring dwarfs, specifically those too faint to be observed by the SDSS, we expect this contamination to be low due to the overall isolation of these pairs. For the TNT nonisolated pairs, however, the resulting \HI\ masses may be upper limits since their nonisolation criterion does not exclude the possibility of neighboring galaxies falling within the beam. 

Our observing strategy was to perform an intial on/off pair with three and five minute on source integration times using the Arecibo L-Band Wide spectrometer and the GBT Spectrometer, respectively. If the target was easily detected in the first on/off pair, no further integration was done. Otherwise, each target was observed until an rms of $\sim$0.6 and 1.7 mJy was reached for Arecibo and the GBT, respectively. These noise levels translate to upper limits of M$_{HI} \lesssim 3 \times$ 10$^8$M$_{\odot}$ and M$_{HI} \lesssim$ 10$^9$M$_{\odot}$ at the median distance for our pairs of 125 Mpc and assuming a velocity width of 150 \kms\ and a signal to noise ratio of five. 
Flux calibration for the GBT was done by observing a flux calibrator (usually a quasar) every 3-4 hours to determine the L-band aperture efficiency which ranged from 0.58 to 0.75. Flux calibration for Arecibo was done by taking spectra of a built-in noise diode of known system temperature T$_{sys}$ after each on/off pair. Spectra for both the GBT and Arecibo samples were Hanning smoothed to a velocity resolution of 8 \kms.

Eleven pairs were observed with both the Arecibo and the GBT single dish telescopes to test for consistency. Eight are in agreement given the relative sensitivities (i.e. either the \HI\ line spectra agree within 5\% or the more sensitive Arecibo observations detect a signal that is within the noise of the GBT spectrum). The remaining three pairs appear to have more flux in the GBT spectrum which results in \HI\ fluxes that are $\sim$20\% higher than measured with Arecibo. Given the extent of the Magellanic Stream \citep[$>$100 kpc;][]{putmanstream}, a similarly extended diffuse feature may be detected within the extent of the GBT beam (i.e. 100-260 kpc at the distances of these three pairs) and not within the Arecibo beam ($<$ 100 kpc). 

\begin{figure*}
\begin{sideways}
\begin{minipage}[t]{8.5in}
\centering
\hspace{-0.8cm}
\includegraphics[height=1.8in,width=1.8in]{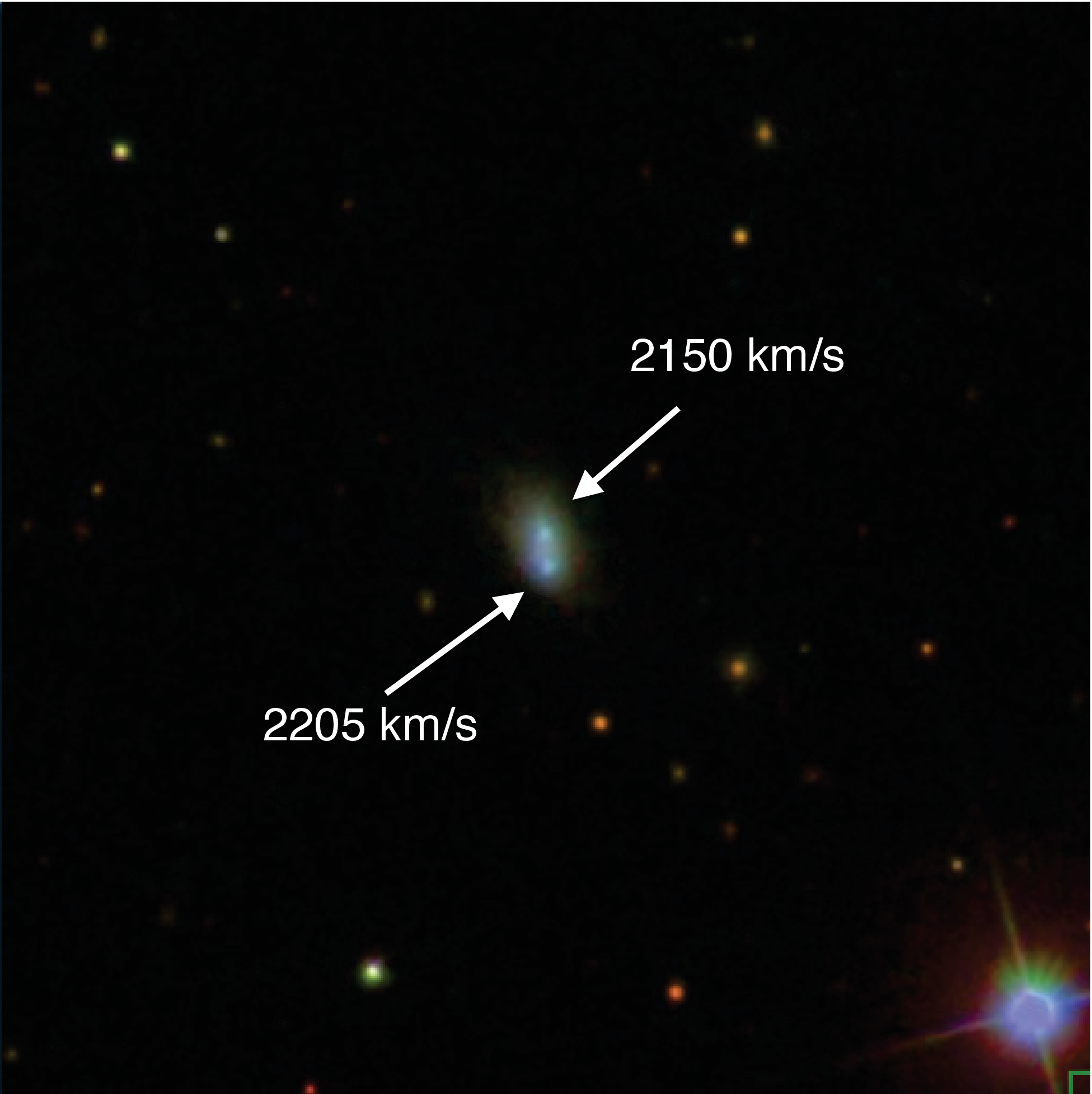}
\hspace{+.2cm}
\includegraphics[height=1.8in,width=1.8in]{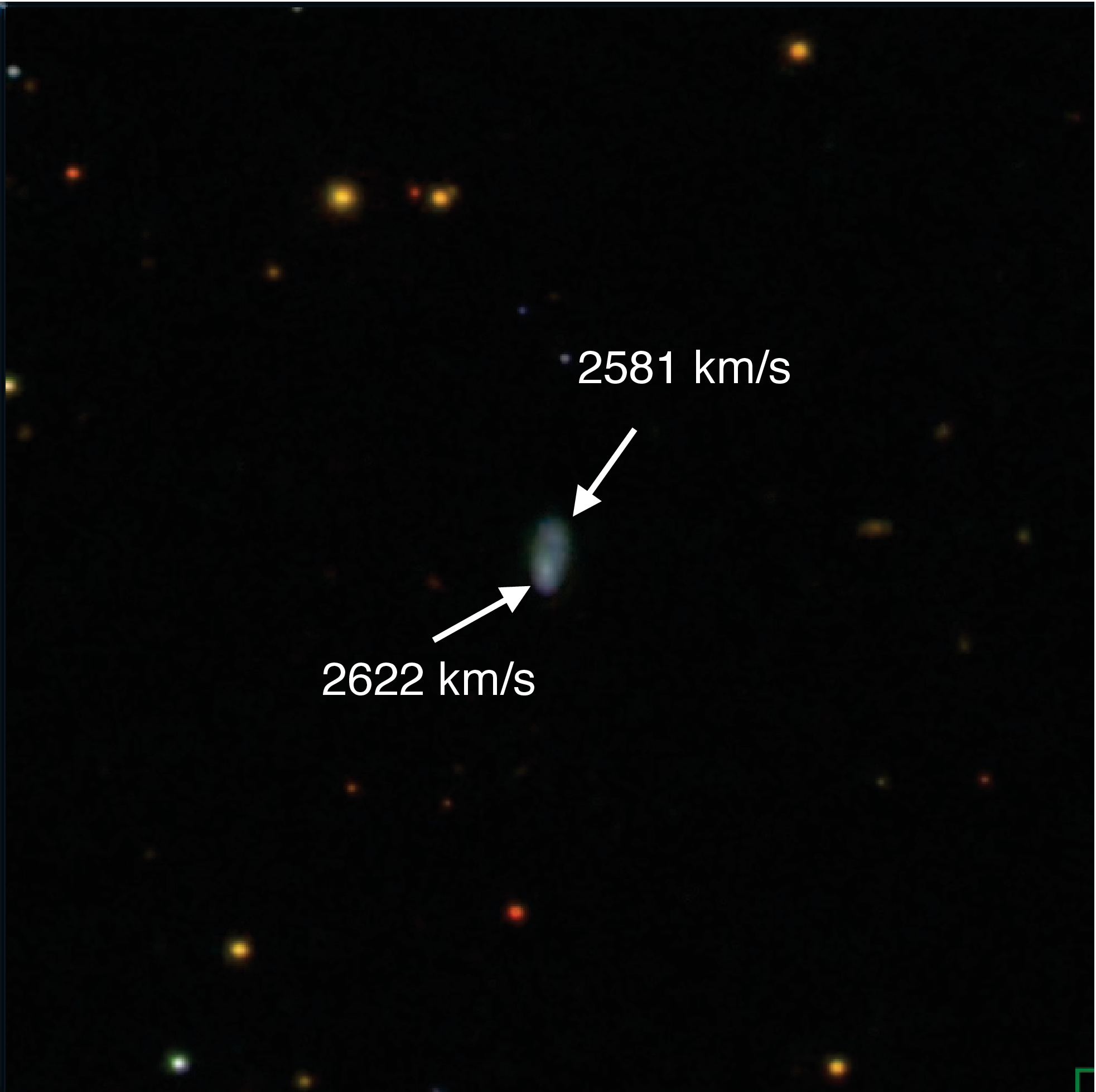}
\hspace{+.2cm}
\includegraphics[height=1.8in,width=1.8in]{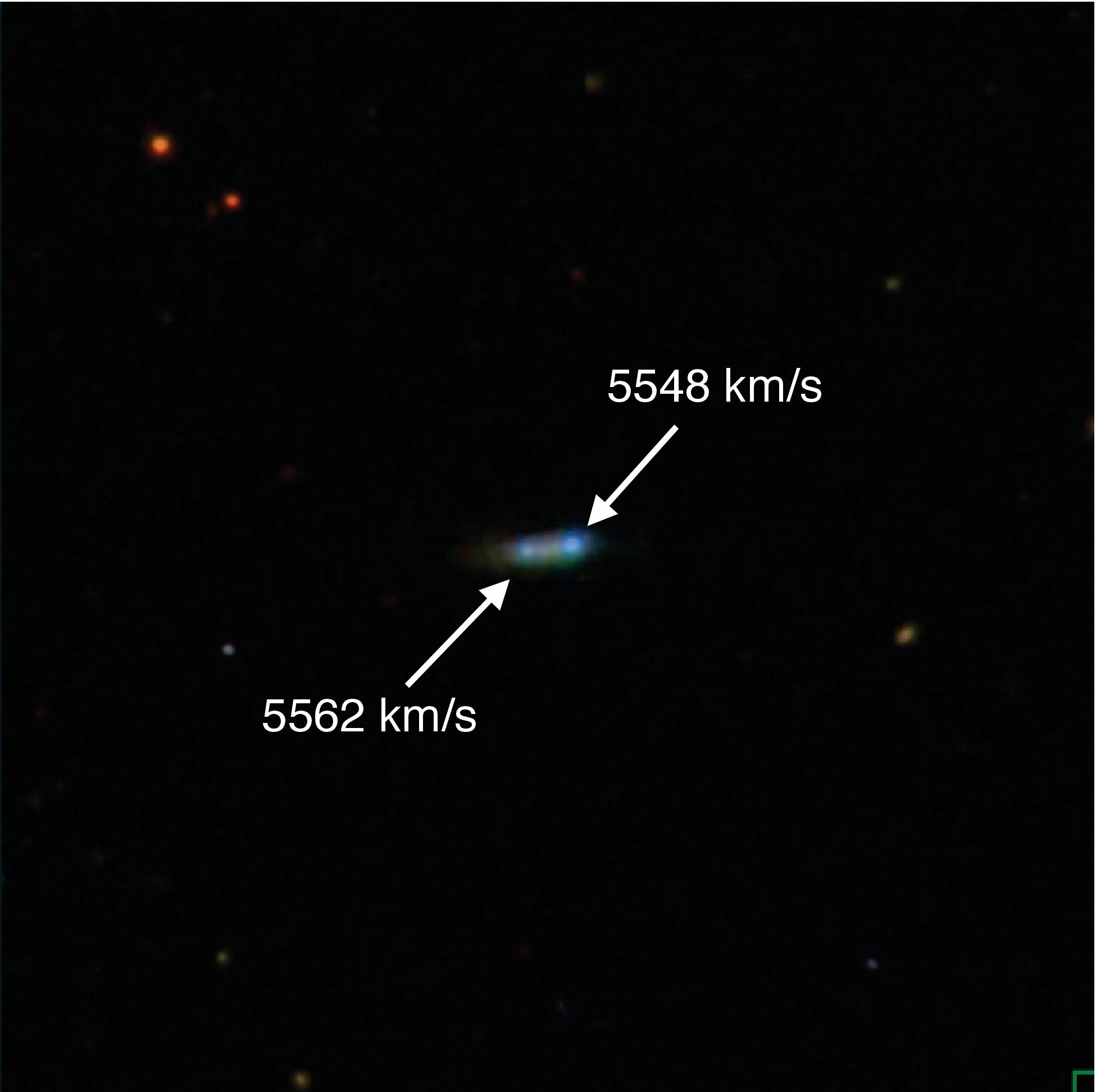}
\hspace{+.2cm}
\includegraphics[height=1.8in,width=1.8in]{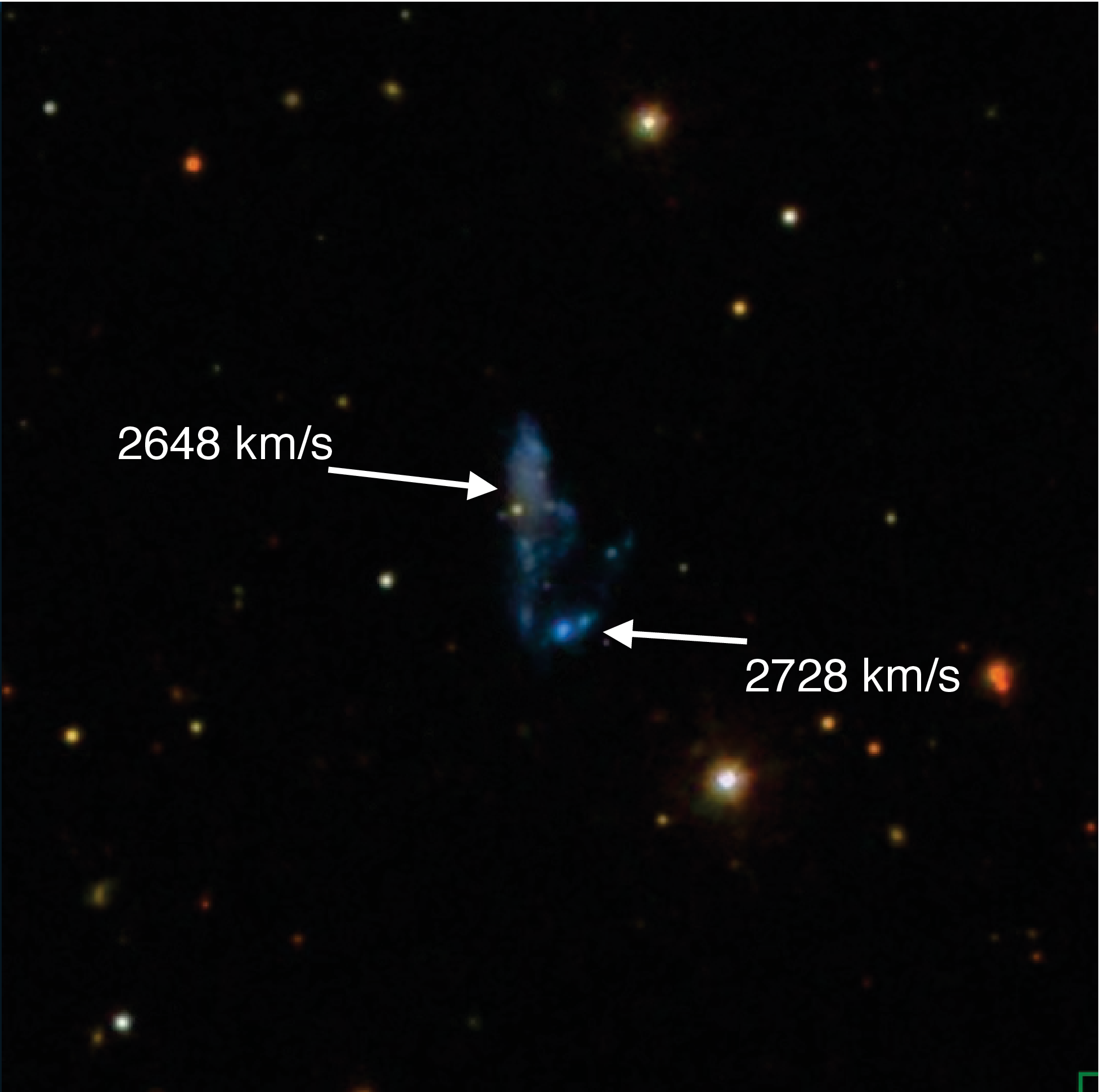}
%\hspace{-2cm}
\end{minipage}
%\hfill
\end{sideways}
\begin{sideways}
\begin{minipage}[t]{8.5in}
\vspace{-.3cm}
\centering
\hspace{-1.5cm}
\includegraphics[width=2.2in,height=1.4in,clip]{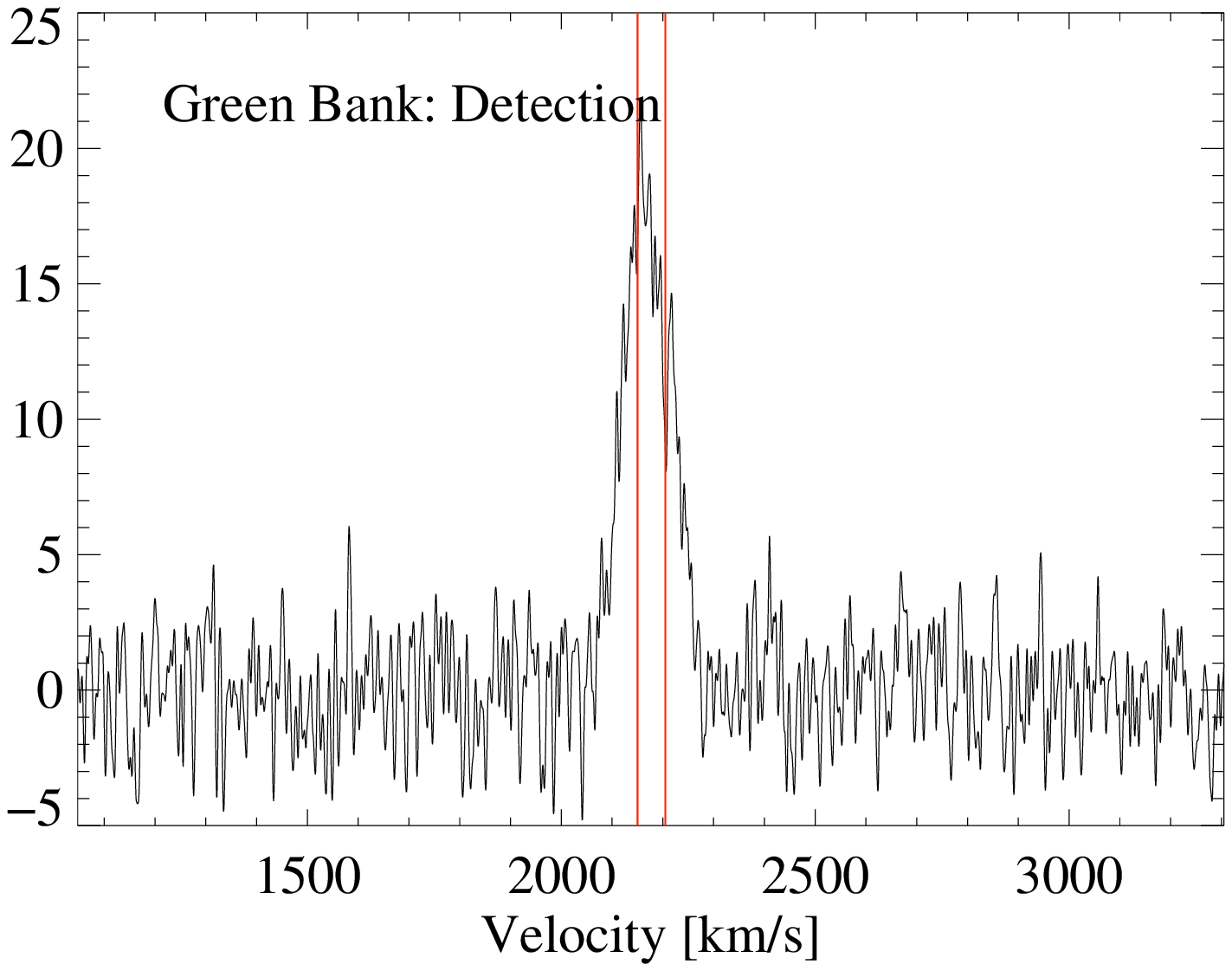}
\hspace{-.8cm}
\includegraphics[width=2.2in,height=1.4in,clip]{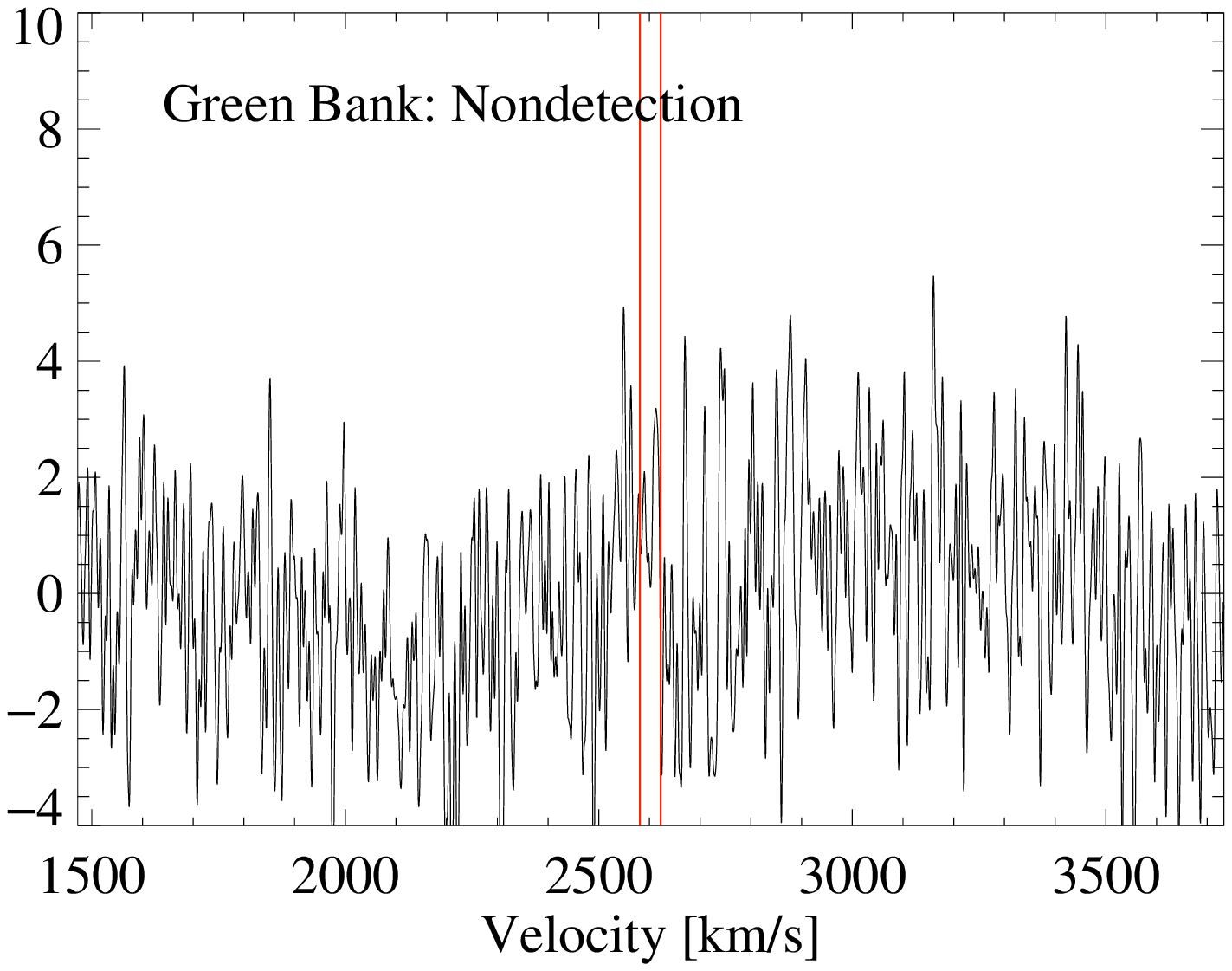}
\hspace{-.8cm}
\includegraphics[width=2.2in,height=1.4in,clip]{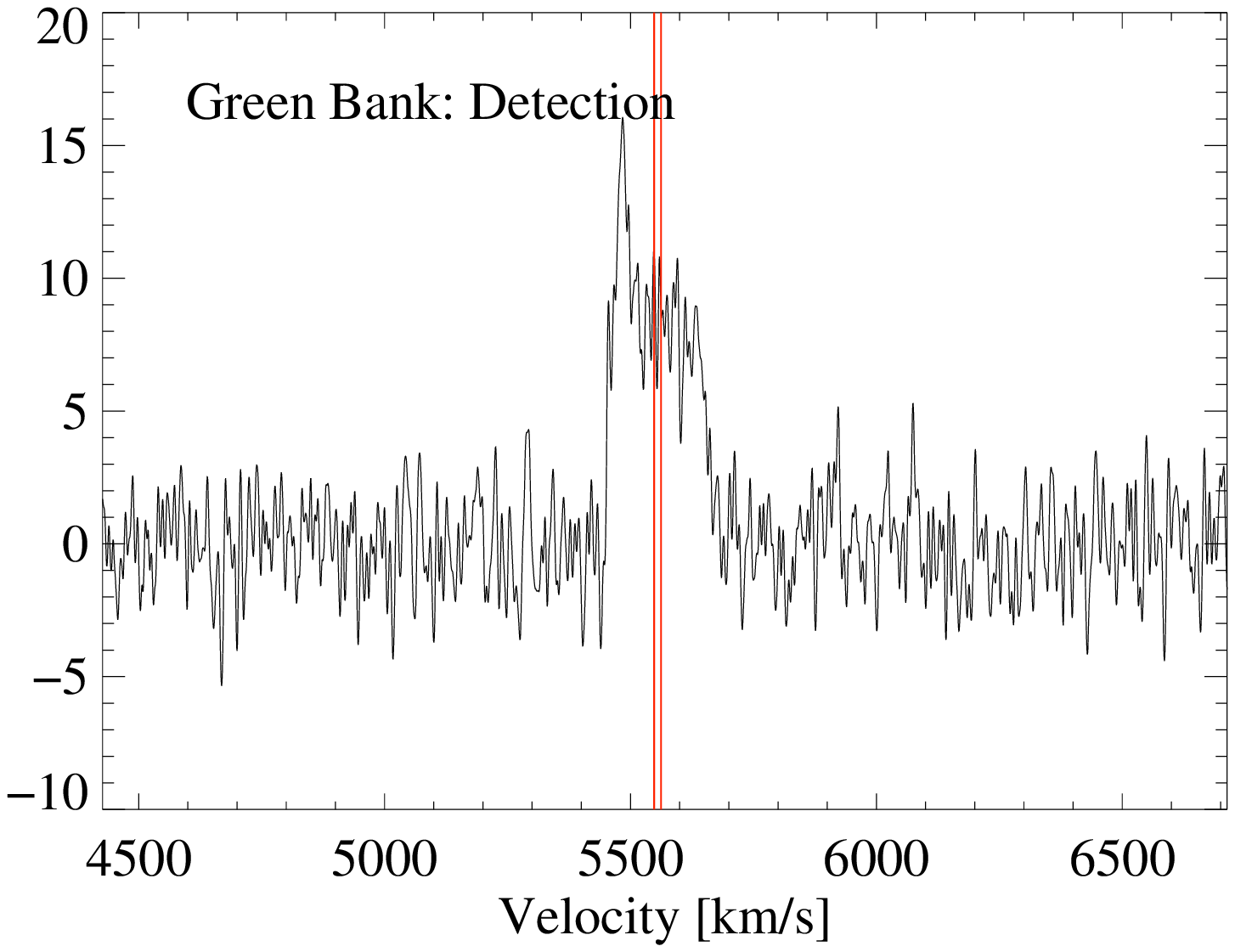}
\hspace{-.8cm}
\includegraphics[width=2.2in,height=1.4in,clip]{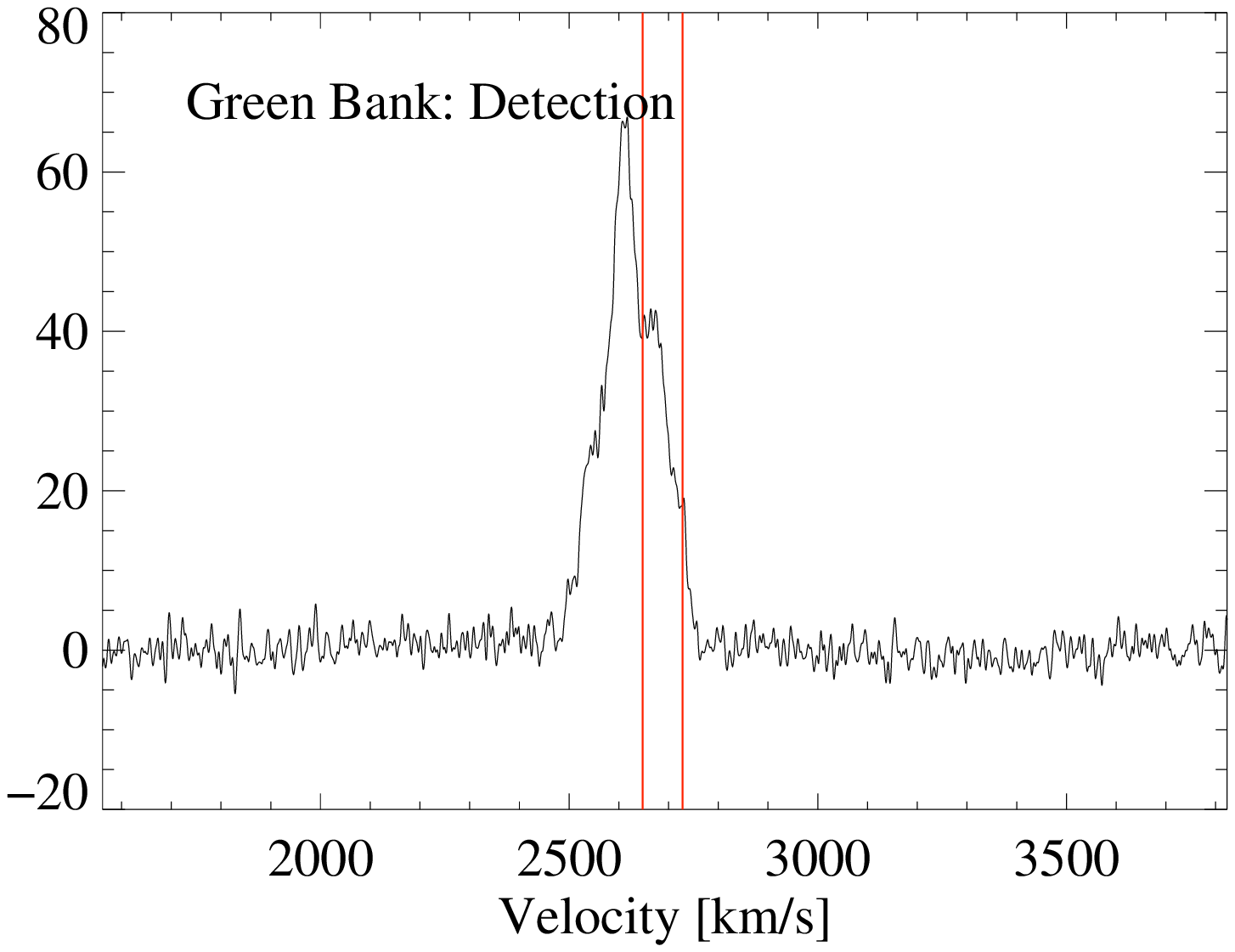}
\end{minipage}
%\hfill
\end{sideways}
\begin{sideways}
\begin{minipage}[t]{8.5in}
\vspace{-0.2cm}
\centering
\hspace{-0.8cm}
\includegraphics[height=1.8in,width=1.8in]{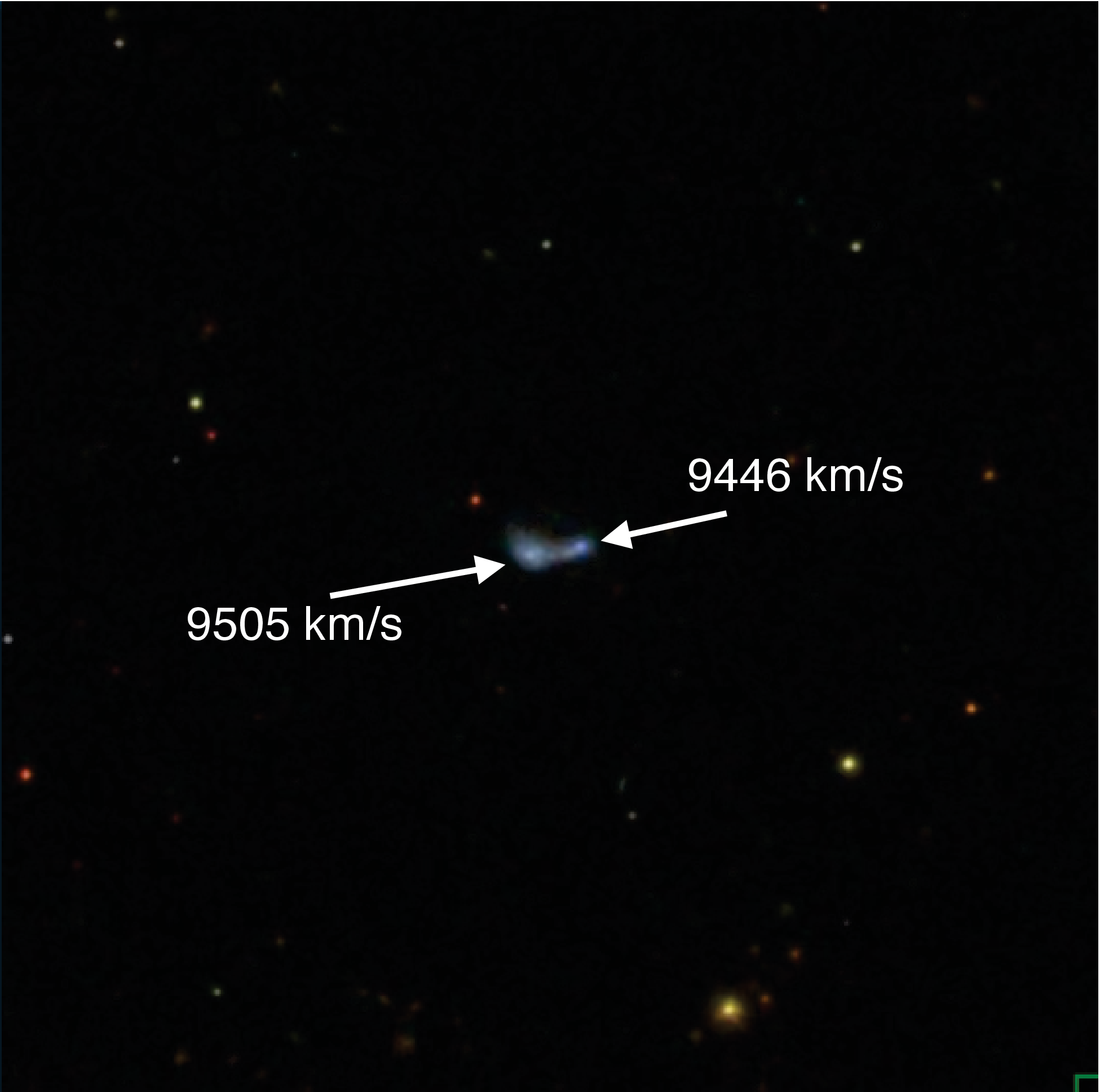}
\hspace{+.2cm}
\includegraphics[height=1.8in,width=1.8in]{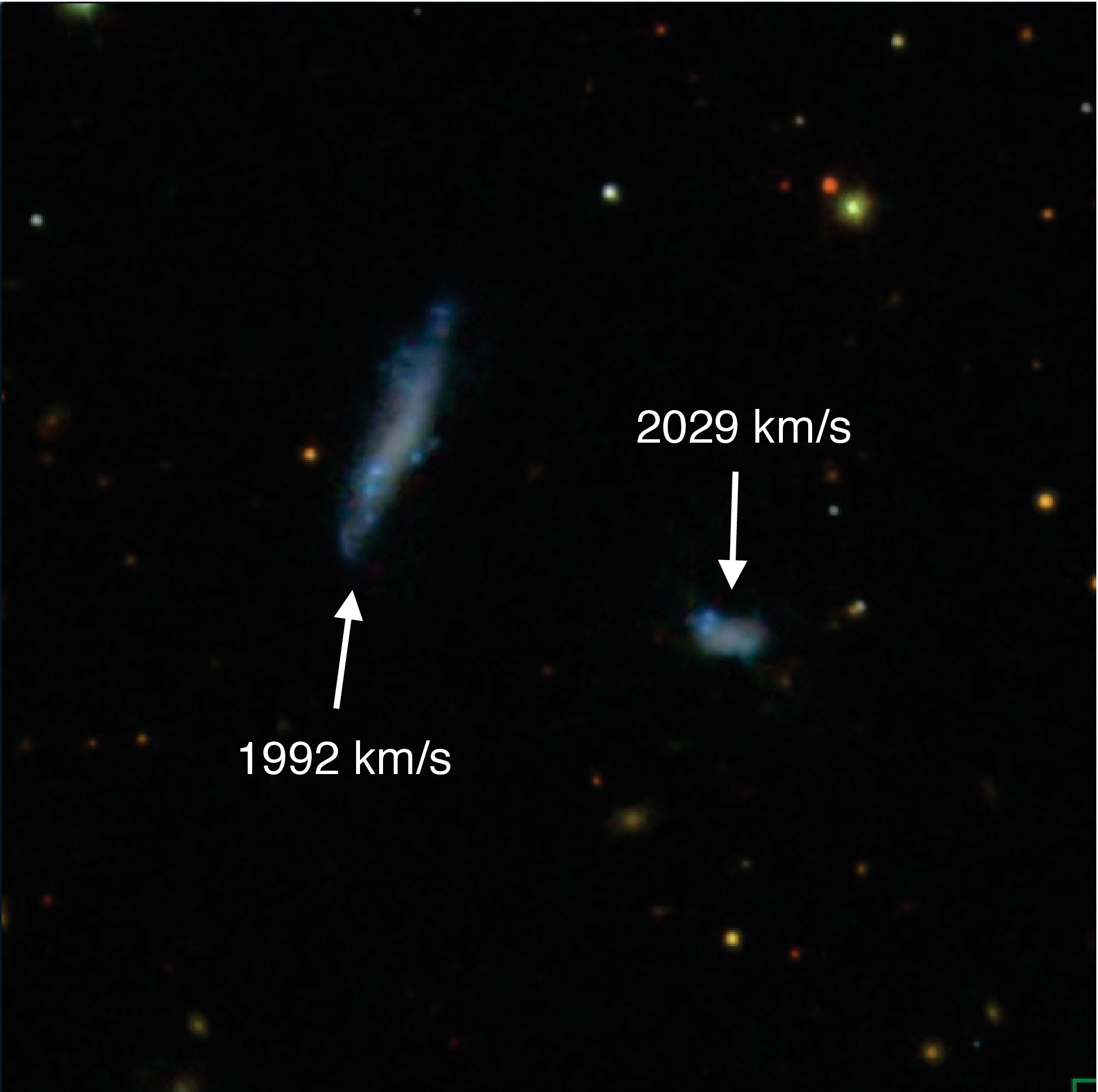}
\hspace{+.2cm}
\includegraphics[height=1.8in,width=1.8in]{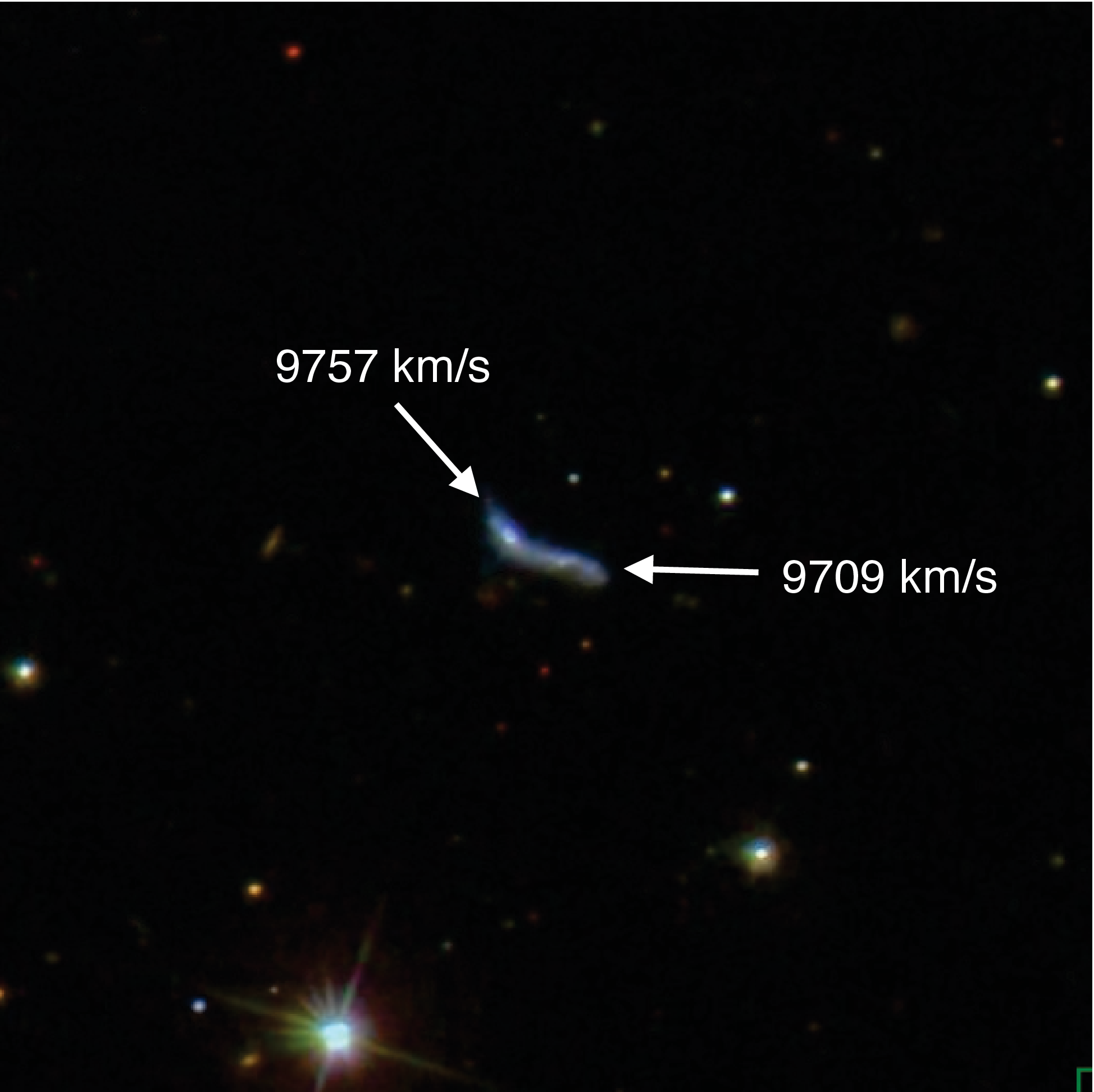}
\hspace{+.2cm}
\includegraphics[height=1.8in,width=1.8in]{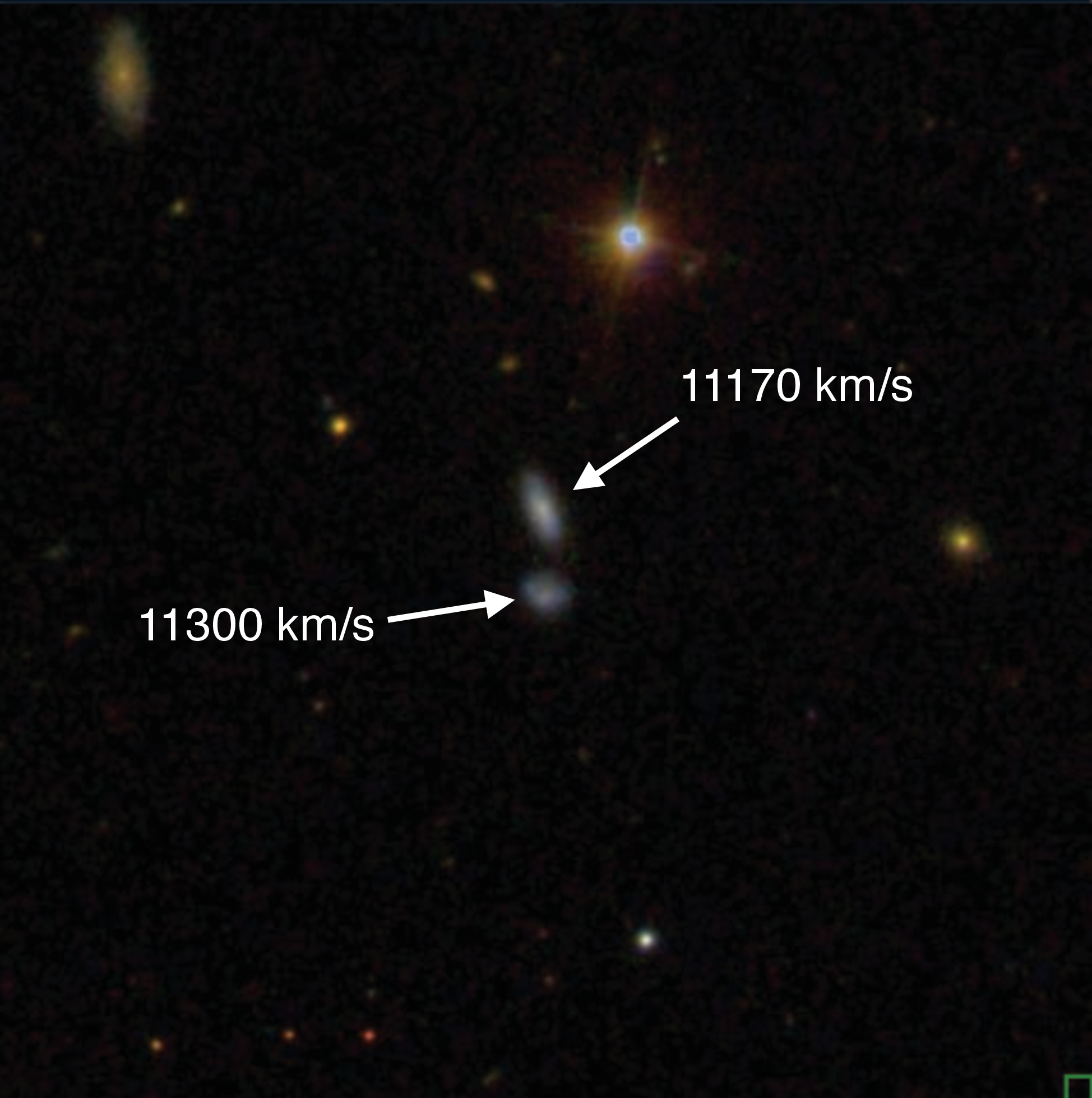}
%\hspace{-2cm}
\end{minipage}
%\hfill
\end{sideways}
\begin{sideways}
\begin{minipage}[t]{8.5in}
\vspace{-.3cm}
\centering 
\hspace{-1.5cm}
\includegraphics[width=2.2in,height=1.4in,clip]{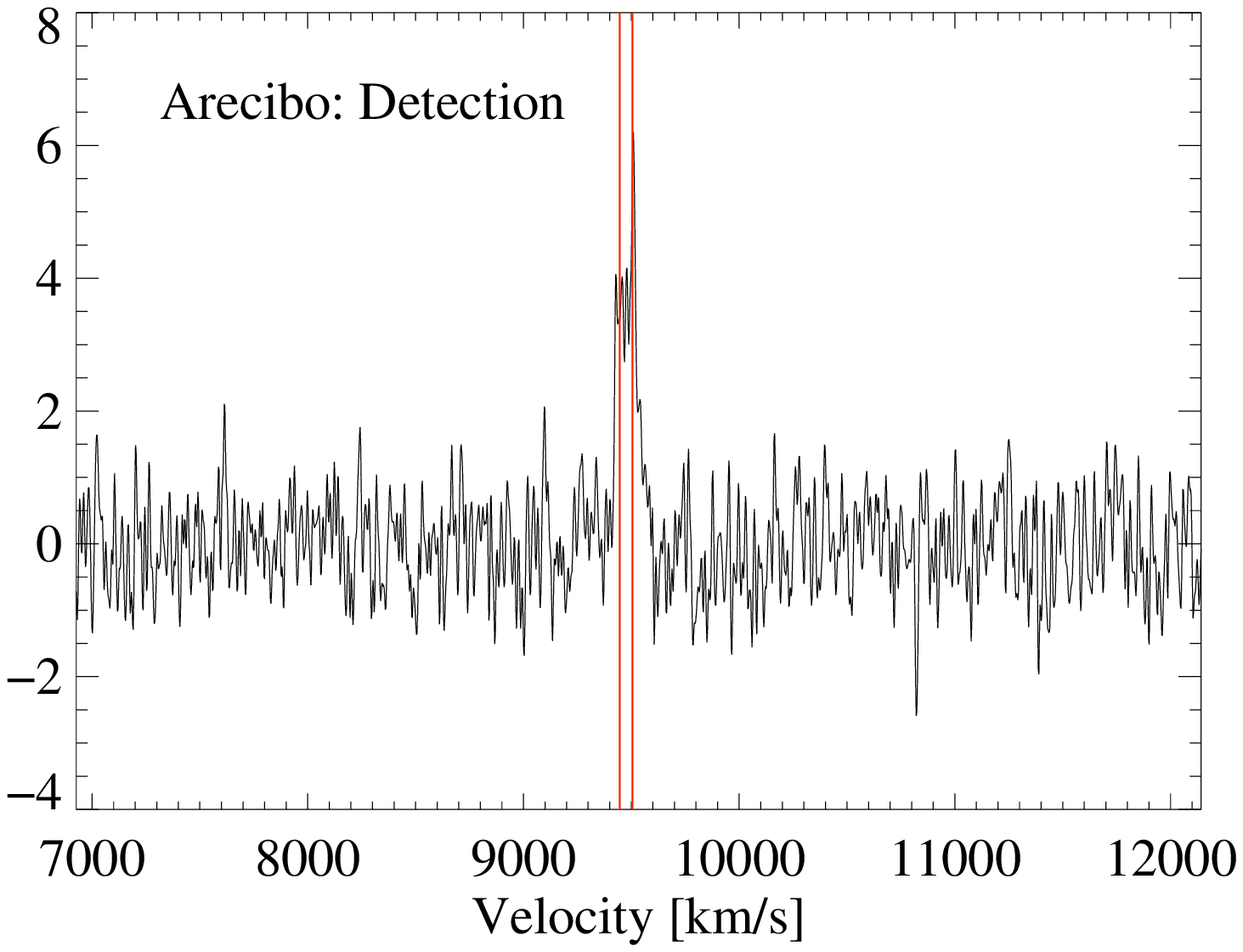}
\hspace{-.8cm}
\includegraphics[width=2.2in,height=1.4in,clip]{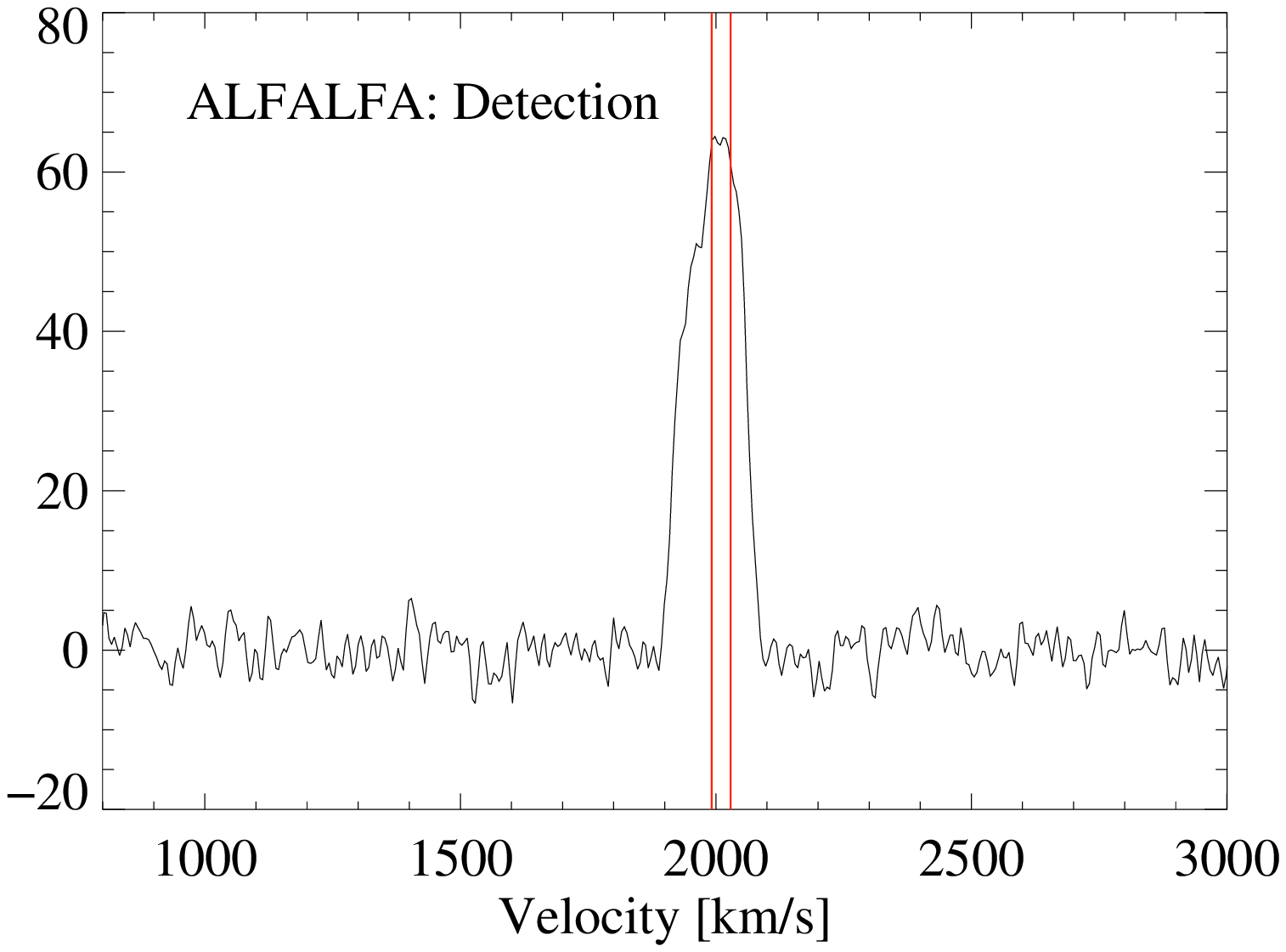}
\hspace{-.8cm}
\includegraphics[width=2.2in,height=1.4in,clip]{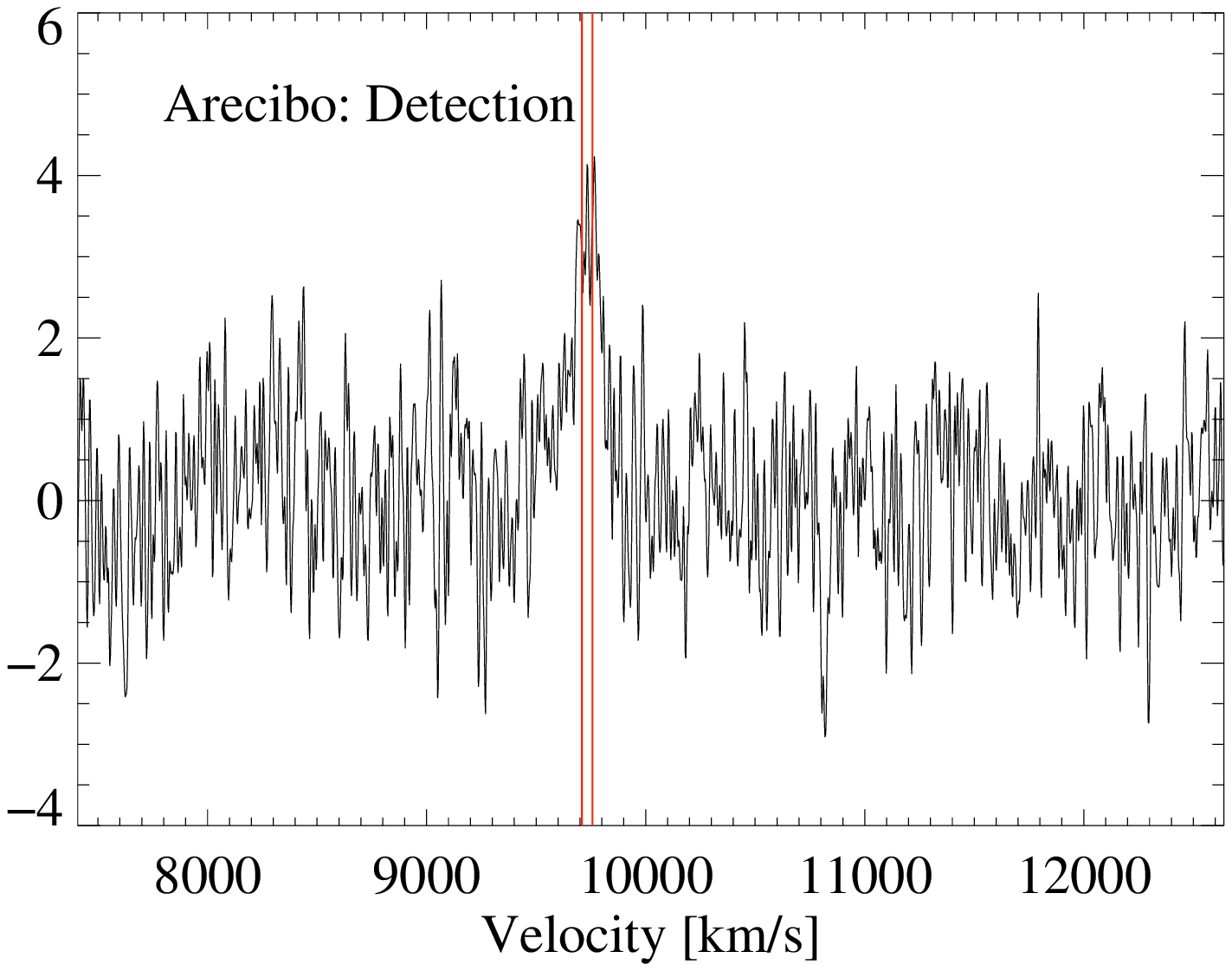}
\hspace{-.8cm}
\includegraphics[width=2.2in,height=1.4in,clip]{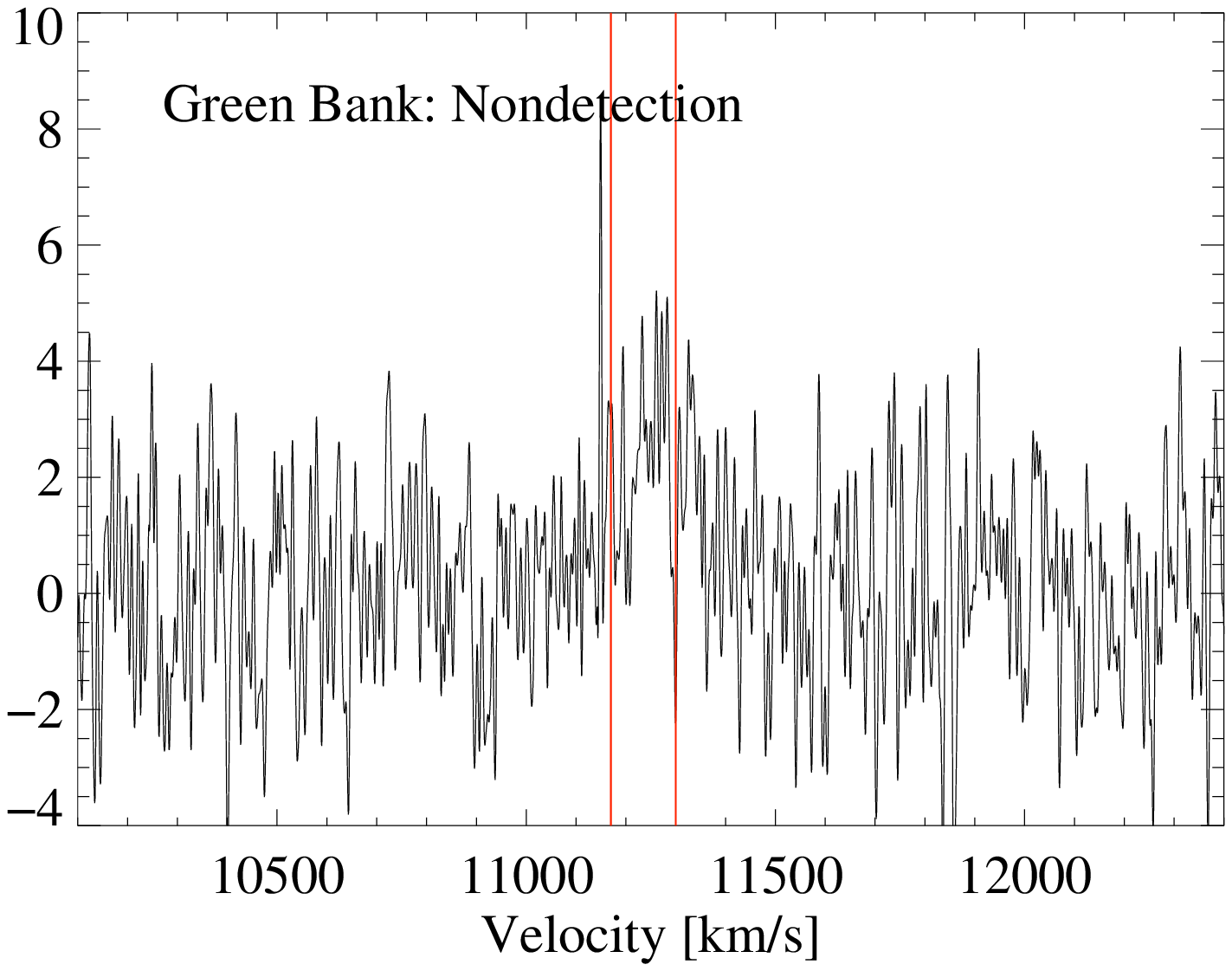}
\end{minipage}
%\hfill
\end{sideways}
%\hspace{-1cm}
\begin{sideways}
\begin{minipage}[t]{8.5in}
%\vspace{-0.1cm}
\centering
\caption{Broadband SDSS ugriz images and new single-dish Arecibo and GBT \HI\ line spectra for the first 8 TNT isolated dwarf pairs (in order of projected radial separation). Images are 200\arcsec$\times$200\arcsec. Optical velocities for each pair member are labeled and noted as red vertical lines on each spectrum. Spectra are given in units of km/s and mJy. \HI\ spectra marked ALFALFA come from the Cornell Digital \HI\ Archive. The remaining 52 TNT isolated pairs are shown in the appendix.}\label{atlas}
\end{minipage}
\end{sideways}
\end{figure*}

In Figure \ref{atlas}, we show the observed \HI\ line spectrum along with the SDSS broadband image for each pair. We mark the velocity of the optical redshifts of each pair member (as measured from the SDSS spectra) as red lines on each spectrum and with labels in the optical image. We also note which single-dish telescope was used to produce each spectrum and whether or not we consider it a detection. For the 10 pairs for which \HI\ spectra were already available from ALFALFA, the spectra shown here are as they appear in the Cornell Digital \HI\ Archive.

Assuming the \HI\ is optically thin, we determine the \HI\ mass using the formula:
\begin{equation}
M_{HI}/M_{\odot} = 2.356 \times 10^5D^2_{Mpc} S_{int}
\end{equation}
\noindent where S$_{int}$ is the integrated flux density in Jy \kms\
and D$_{Mpc}$ is the average distance to the two pair members in
Mpc. For pairs with an average velocity $\langle$v$\rangle >$ 3000 \kms, we set the
distance to D$_{Mpc} = \langle$v$\rangle$/H$_0$ and use H$_0 =$ 70
\kms ~Mpc$^{-1}$ for the Hubble constant \citep{hnot}. For closer pairs, where
peculiar velocities may compete with those due to Hubble flow, we use
the peculiar velocity model of \cite{tonryflow} to determine distances. For the 10 pairs with spectra in the Cornell Digital \HI\ Archive, we compare our re-analysis of the spectra with the published results in the Arecibo-based ALFALFA $\alpha$40 catalog \citep{a40}. We find our resulting integrated \HI\ fluxes and line widths agree within 5\%.

\section{Enhanced Star Formation and Starbursts in Paired Dwarfs}\label{sfr}
Star formation in nearby, paired $massive$ galaxies is observed to be enhanced over that of unpaired galaxies of similar stellar mass \citep[median galaxy mass M$_* =$ 10$^{10}$ M$_{\odot}$;][]{patton13} out to pair separations of $\sim$100 kpc, which is roughly 40\% of the typical virial radius, R$_{vir}$, of a Milky Way type galaxy. The observed enhancement increases as the separation between the pair members decreases and is the most significant (a factor of $\sim$2.5) at separations R$_{sep} < $25 kpc ($\sim$10\% of R$_{vir}$). This enhancement is further observed to be even larger in massive post-merger galaxies than for the closest (i.e. late stage) pairs \citep{ellison13}. Numerical simulations predict bursts of star formation are triggered by each pericentric passage of two interacting massive galaxies. The resulting SFR is enhanced relative to isolated analogs, even as the two galaxies move to wider separations. \citep[][and references therein]{cox08,hopkins08,patton13}.

Such observational and theoretical studies of galaxy pairs typically focus on galaxies with M$_* > $10$^9$M$_{\odot}$. In the lower mass dwarf regime, the long range gravitational forces may not be sufficient to drive global enhancements in SFR. However, there is evidence across a large range of stellar masses (10$^7 < $M$_*$/M$_{\odot} < $ 10$^{12}$), including dwarf mass scales, that galaxies identified as starbursts show some indication of a past merger or interaction via tails, bridges, or double cores \citep{luo}. These interaction-triggered starbursts may eventually produce the so far unexplained, unusually high star formation rates and compact surface brightness profiles in nearby blue compact dwarf galaxies \citep[BCDs; e.g.,][]{lelli14,lelliSB1,lelliSB2,janoBCD}. Whether or not BCDs are dwarf-dwarf merger remnants, or if such starbursts can be induced at earlier phases of the dwarf-dwarf interaction sequence, is unclear. The secondary member in the interacting pair NGC4485/4490 is known to be starbursting \citep{11HUGS}, but the larger sample of dwarf pairs covering a range of separations provided by TNT is necessary to determine if this system is unique.

For the remainder of this section, we explore whether dwarf pairs exhibit trends in SFR as a function of pair separation and determine whether dwarf-dwarf encounters are capable of inducing starbursts in the stages of the encounter prior to final coalescence.

\subsection{Enhanced Star Formation in Paired Dwarfs}

\begin{figure*}
\begin{center}
\includegraphics[height=2.3in,width=2.9in]{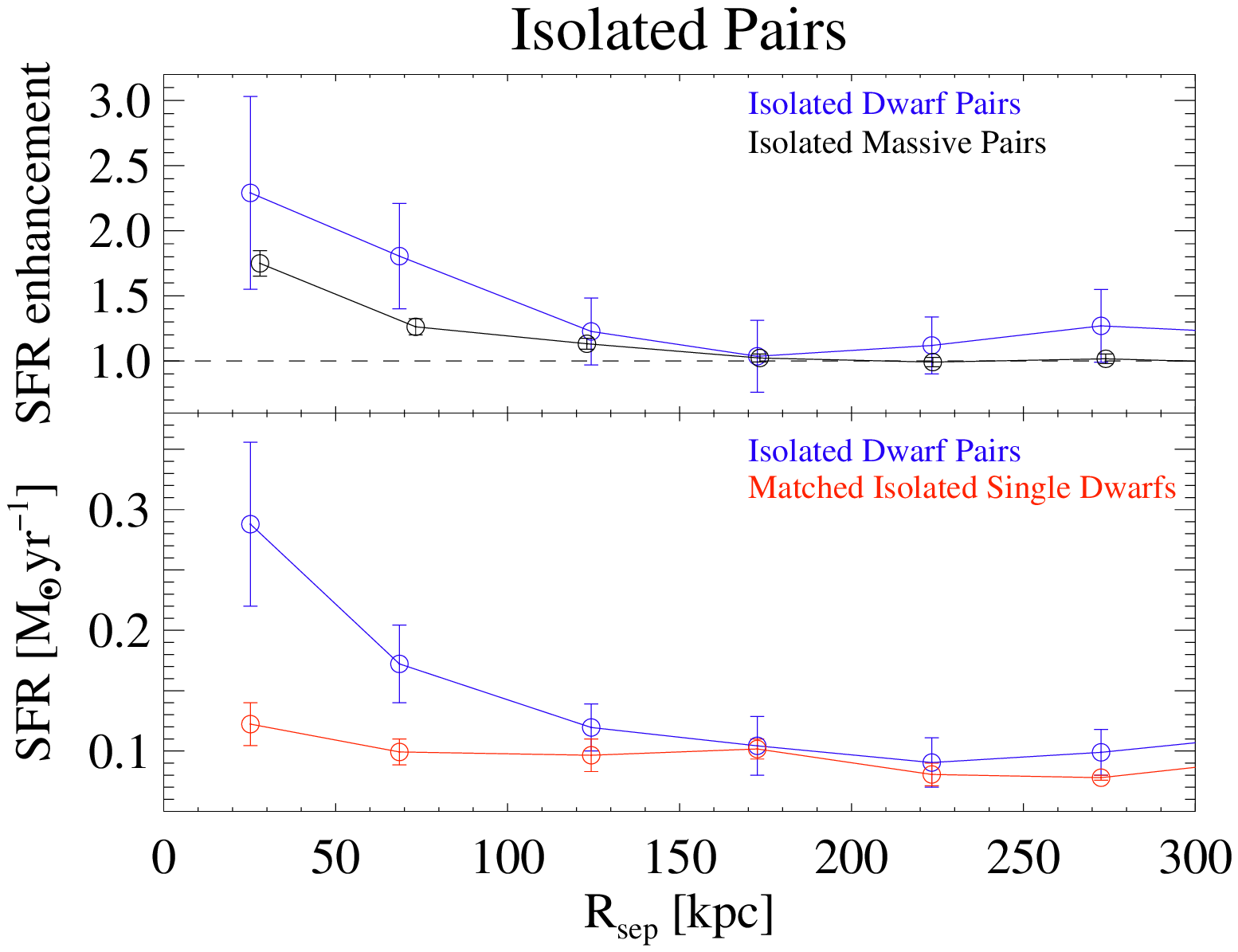}
\includegraphics[height=2.3in,width=2.9in]{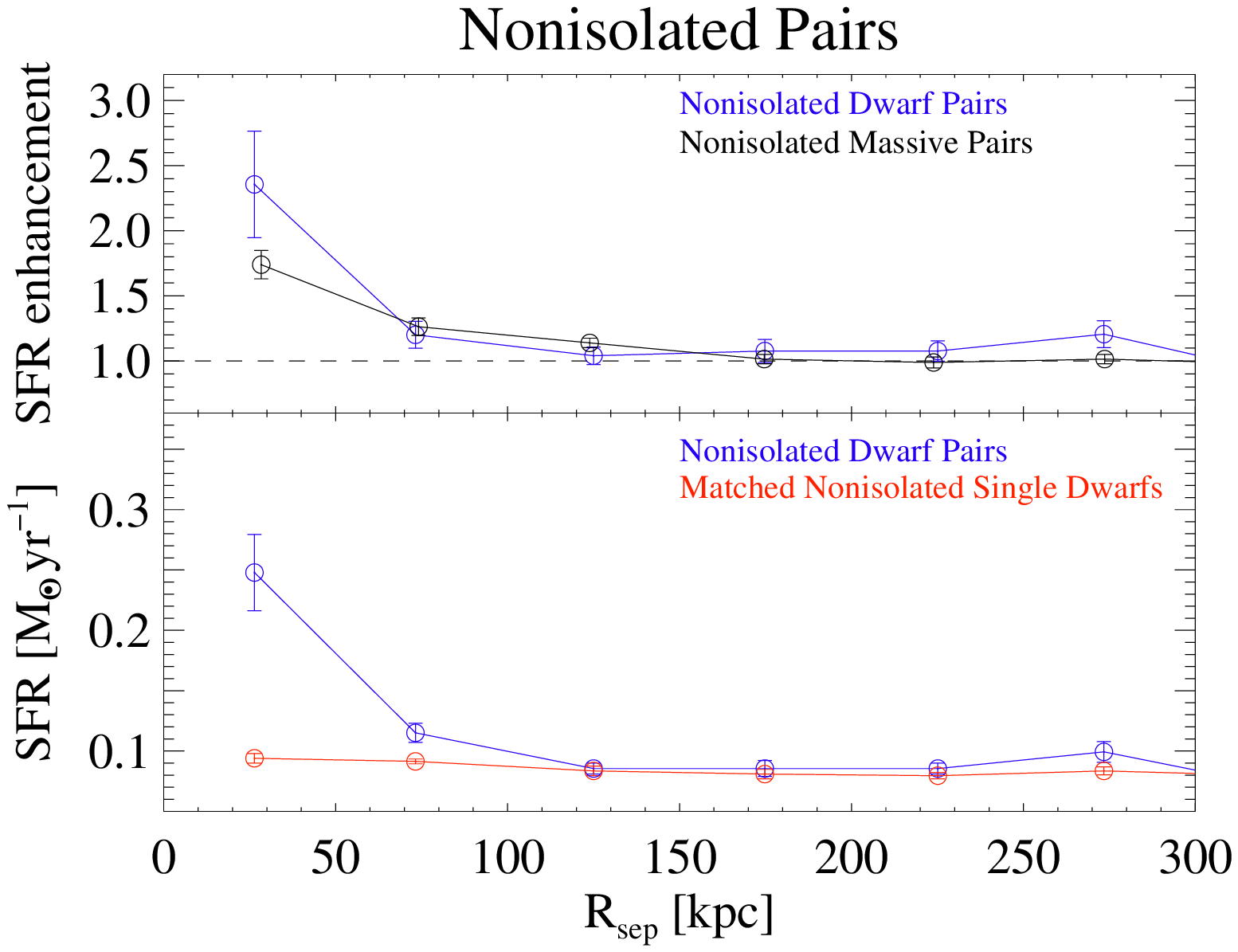}
\caption{Star formation enhancement for isolated (left) and nonisolated (right) paired dwarfs. Lower panel: The mean SFR for the dwarf galaxy pair members from the main TNT sample (lowest radial bin) and extended to larger radial pair separations (blue) and their associated matched isolated single dwarfs (control galaxies matched in M$_*$, redshift, and local density and isolation; red) is plotted versus projected separation (R$_{sep}$) for the paired dwarfs. Upper panel: Mean SFR enhancement (ratio of pair member SFR to control SFR) is plotted versus R$_{sep}$ for dwarf pairs (blue) and massive galaxy pairs \citep[black;][]{patton13}, with the dashed horizontal line denoting zero enhancement. Data is averaged in radial bins of 50 kpc and points are plotted in the middle of the radial bin they represent. All error bars show the standard error in the mean.
\label{sfrenhance}}
\end{center}
\end{figure*}

In Figure \ref{sfrenhance}, we show that star formation in both the TNT isolated and nonisolated dwarf pair samples is enhanced relative to their matched single dwarf control samples (defined in Section \ref{controls}). The lower panels show the average SFR for individual pair members from our extended pair sample in bins of 50-kpc separation as a function of projected radial separation (blue line) for isolated (left) and nonisolated (right) dwarf pairs. Following \cite{patton13}, we apply a weighting scheme to account for incompleteness due to fiber collisions in these averages. The average SFRs for unpaired dwarfs in bins of 50-kpc separation, as a function of the projected radial separations of the paired dwarfs to which they are matched, are plotted in red for isolated (left) and nonisolated (right) unpaired dwarfs. In both the isolated and nonisolated cases, the ratios of the blue and red lines (the average SFR of paired dwarfs vs the average SFR of unpaired dwarfs, i.e. the SFR enhancement) are shown in the top panels (blue lines). The SFR enhancement observed for massive paired over unpaired galaxies \citep[determined using a similar method, see][]{patton13} is shown for both isolated and nonisolated massive galaxy pairs (black lines). The enhancement seen for the dwarf pairs appears more significant by a factor of 1.3 than that observed for more massive galaxies. However, the results for the two pair samples are still consistent within their given errors, and so larger statistics, particularly for the dwarf pairs, are needed to lower the uncertainties and make this result conclusive. 

The enhancement in isolated pairs is observed as far as separations of $\sim$100 kpc for both dwarfs and massive galaxies. Since 100 kpc translates to roughly a virial radius for the dwarfs in our sample but only $\sim$40\% of R$_{vir}$ for more massive galaxies (R$_{vir}\sim$250 kpc), star formation induced during prior pericenter passages may have longer lifetimes and be more important throughout the dwarf-dwarf interaction rather than in more massive galaxy mergers.

An enhancement of similar strength to that observed for the isolated pairs is also observed for the nonisolated pairs, but does not extend out past R$_{sep} \sim$ 50 kpc. Thus, regardless of environmental factors like a nearby massive neighbor, close encounters between pairs of dwarfs will increase their SFR. This effect may explain why the SFR of both the LMC and SMC (R$_{sep}\sim$ 15 kpc) have been increasing over the past few Gyr \citep{harrisLMC,weiszLMC} despite their proximity to the Milky Way, which is otherwise expected to cause a plateau or decline in their SFRs \citep{wetzelquench}.

\subsection{Starburst Fraction in Paired Dwarfs}

To determine whether dwarf-dwarf interactions are powerful enough to trigger starbursts, as is observed in massive galaxy mergers, we determine the starburst fraction of the TNT sample in three different ways. First, we define an individual dwarf pair member as a starburst if its SFR is more than five times the average SFR for its matched controls. Second, since the number of controls found to match a given pair member varies with each dwarf, we also determine the starburst fraction by defining a starburst as a dwarf with a SFR of more than five times the average SFR of all star-forming galaxies of similar stellar mass in SDSS \citep[i.e. the star-forming main sequence, SF MS, as parameterized in][]{luo}. Finally, we also determine the starburst fraction using a simple cut in H$\alpha$ equivalent width (H$\alpha$ EQW $>$ 100\AA) to remove the need for a comparison sample and for ease with comparison to previous studies. H$\alpha$ equivalent widths measured from the SDSS spectra, and not corrected for internal extinction, are available for all of our samples via the MPA-JHU DR7 data release\footnote{http://www.mpa-garching.mpg.de/SDSS/DR7/} \citep[i.e.][]{jarle04}. 

The resulting starburst fractions using each of these three definitions of a starburst, as well as the quenched fraction (i.e. fraction with H$\alpha$ EQW $<$ 2\AA) are presented for the isolated pair sample, the nonisolated pair sample, the matched isolated single dwarfs, and the matched nonisolated single dwarfs in Table 1. For all three starburst metrics, the isolated pair members show a higher fraction of starbursts than either single dwarf sample, suggesting that dwarf-dwarf interactions are a key mode of triggering dwarf starbursts. The starburst fractions observed for our single dwarf samples agree well with the results of \cite{11HUGS}, which found 6\% of dwarfs within the local volume (D $<$ 11 Mpc) to be starbursts.

\begin{table*}[htbp]
\begin{center}
{\bf{Table 1: TNT Starburst and Quenched Fractions}}\\
\begin{tabular}{l c c c c}
\hline
Sample & SB Frac & SB Frac & SB Frac & Quenched\\
& (controls) & (SF MS) & (H$\alpha$) & (H$\alpha$)\\
\hline
Isolated Pairs & 15\% ($\pm$4\%) & 16\% ($\pm$3\%) & 20\% ($\pm$2\%) & 0\% ($\pm$0\%) \\
\hline
Nonisolated Pairs & 12\% ($\pm$2\%) & 5\%  ($\pm$2\%) & 20\% ($\pm$2\%) & 8\% ($\pm$3\%) \\
\hline
Isolated Singles & --- & 3\% ($\pm$1\%) & 6\% ($\pm$2\%) & 0\% ($\pm$0.4\%)\\
\hline
Nonisolated Singles & --- & 3\% ($\pm$1\%) & 8\% ($\pm$1\%) & 5\% ($\pm$2\%) \\
\hline
\end{tabular}
\end{center}
\indent Starburst and quenched fractions for the TNT samples using three different metrics. Column (1): TNT Sample Name, Column (2): starburst fraction when defining a starburst with SFR $>$ 5$\times \langle$SFR$\rangle$ where $\langle$SFR$\rangle$ is determined using the matched control samples (matched isolated and nonisolated single dwarfs), Column (3): starburst fraction when defining a starburst with SFR $>$ 5$\times \langle$SFR$\rangle$ where $\langle$SFR$\rangle$ is determined using the parameterization of the star formation main sequence for star-forming galaxies in SDSS from \citep{luo}, Column (4): starburst fraction when defining a starburst by H$\alpha$ EQW $>$100\AA, Column (5): fraction of quenched dwarfs (H$\alpha$ EQW $<$2\AA).
\end{table*}

\begin{figure}[h!]
\begin{center}
\includegraphics[height=2.4in,width=3.5in]{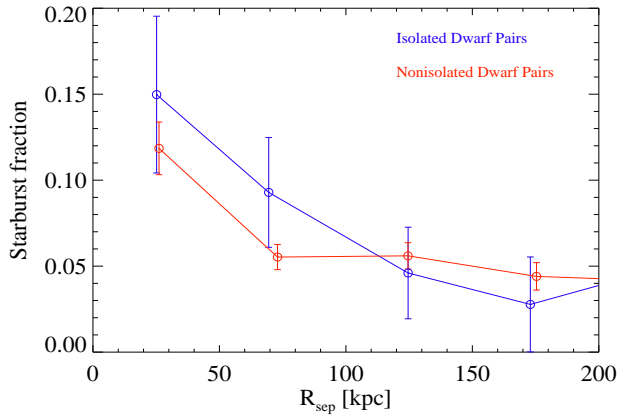}
\caption{Fraction of starbursting dwarfs among the isolated (blue) and nonisolated (red) pair members from the TNT sample extended out to R$_{sep} =$ 200 kpc. Here a dwarf galaxy is considered a starburst if its SFR is more than five times the average SFR of its set of isolated, single ``control'' dwarfs selected to match in redshift, stellar mass, and local density (i.e. Column 2 in Table 1). Both samples show an increase in the starburst fraction at radial separations $<$50 kpc, but the difference in the starburst fraction between close versus more distant pairs is more significant for the nonisolated pair sample.
\label{SBcontrol}}
\end{center}
\end{figure}

We plot in Figure \ref{SBcontrol} for the extended pair sample (i.e. the same sample shown in Figure \ref{sfrenhance}) the fraction of starbursting pair members as a function of radial separation. Here we define a starburst using the first starburst metric in Table 1: each TNT dwarf is compared to the average SFR, $\langle$SFR$\rangle$, of its matched controls, and determined to be a starburst when its SFR $>$ 5 $\times \langle$SFR$\rangle$. Thus the starburst fractions presented in Column (2) of Table 1 represent the results for the first radial bin plotted in Figure \ref{SBcontrol}. Although there is significant scatter due to small number statistics, both the nonisolated and isolated pair members show an increase in the starburst fraction at R$_{sep} <$ 50 kpc to 12\% and 15\% respectively. The nonisolated pair members show a significant difference in starburst fraction for close versus more distant pairs, while the scatter is larger for the isolated pair members. These starburst fractions are lower than the 15-40\% observed for massive galaxy pairs at the same projected separations \citep{ellison13}.

\begin{figure*}
\begin{center}
\includegraphics[height=4.5in,width=5.5in]{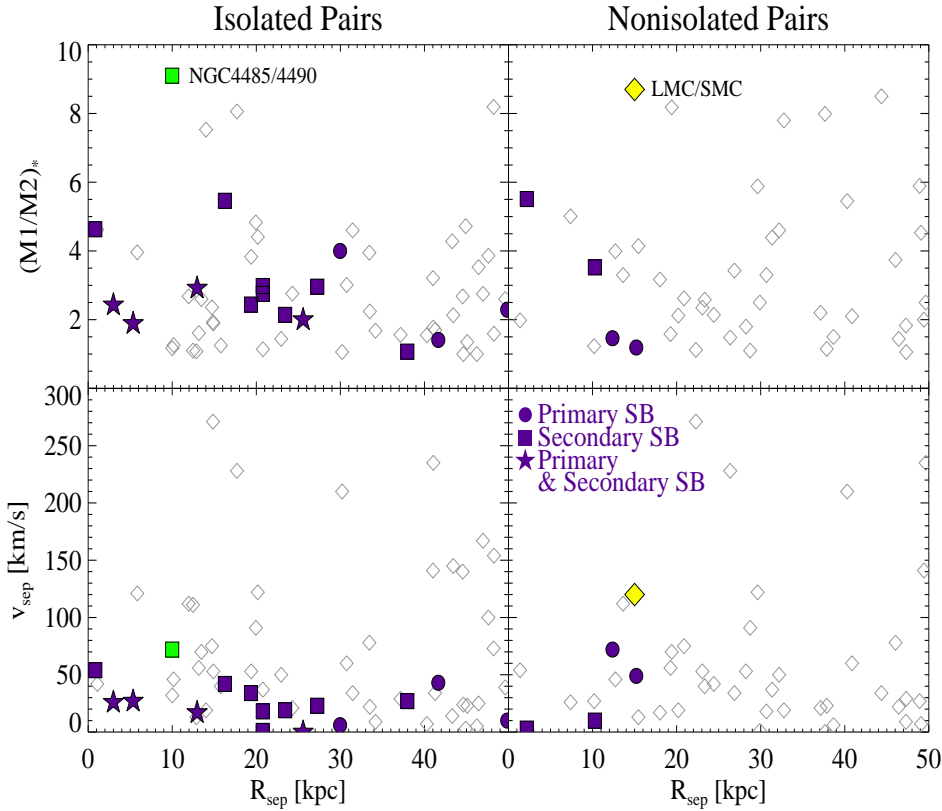}
\caption{Properties of the TNT isolated (left) and nonisolated (right) dwarf pairs that trigger starbursts. For each pair we mark whether the primary (circle), secondary (square), or both (star) pair members are starbursting. Nonstarbursting pairs are shown as gray diamonds. A dwarf pair member is considered starbursting if its SFR is more than five times the average SFR for dwarfs of similar stellar mass, using the star formation main sequence (i.e. stellar mass vs. average SFR relation) of \cite{luo} (see Column 3 of Table 1). Starbursts are more commonly triggered in the secondary pair member and at low relative velocities. In isolated environments, widely separated pairs can host starbursts, but near massive hosts, starbursts are only triggered in those pairs with R$_{sep} <$ 15 kpc. The well-studied interacting pair NGC4485/NGC4490 (green square) and the LMC/SMC (yellow diamond) are marked for reference.
\label{sbproperties}}
\end{center}
\end{figure*}

In Figure \ref{sbproperties}, we explore the pair properties that lead to the starbursts indicated in Figure \ref{SBcontrol} but focus only on the members of the TNT 60 isolated and 44 nonisolated pairs with R$_{sep} <$ 50 kpc. Here we consider a dwarf galaxy to be starbursting if its SFR is more than five times the average SFR for dwarfs of similar stellar mass \citep[using the average SFR as a function of stellar mass determined by][Column (3) of Table 1]{luo}. Note that while 19 members of isolated pairs (16\%) and 4 members of nonisolated pairs (5\%) with available SFRs satisfy this SFR limit, only 3\% of both the isolated and nonisolated single dwarfs would be considered a starburst by this definition.  

In Figure \ref{sbproperties}, we mark whether the primary (circle), secondary (square), or both (star) pair members are starbursting. Nonstarbursting pairs are shown as gray diamonds. In isolated pairs, starbursts appear to be triggered over the full range of radial separations probed by our sample (R$_{sep} < $50 kpc) suggesting that the final coalescence stage of the dwarf-dwarf merger is not required to produce a starbursts but rather they are triggered at earlier stages of the interaction. Starbursts are only triggered at close velocity separations (v$_{sep} <$ 70 \kms) and thus in systems that are more likely to be bound. A similar constraint on v$_{sep}$ is observed for the nonisolated pairs, however starbursts are not observed at separations R$_{sep} > $15 kpc when the pair is in proximity of a massive host galaxy. We do not observe starbursts in either the isolated or nonisolated pairs for stellar mass ratios $(M_1/M_2)_* > 6$, similar to the stellar mass ratios for the well-studied isolated interacting pair NGC4485/NGC4490 (green star) and for the LMC/SMC (yellow star), but, as described in Section \ref{select}, these higher mass ratios are not well sampled by TNT due to the limits of SDSS. Starbursts are more commonly triggered in the secondary (i.e. lower mass) member of the pair, as is shown for more massive galaxy pairs in both observations \citep{woods07} and theory \citep{cox08}. Most pairs for which the primary or both the primary and secondary are starbursting have mass ratios of 1-2.

Although some massive galaxy mergers are shown to end in quiescent systems, likely after having exhausted their fuel for star formation, we find no ``quenched'' dwarfs (i.e. H$\alpha$ EQW $<$ 2\AA) in the TNT isolated pairs sample. There are also no quenched dwarfs among the matched isolated single dwarfs, which is consistent with the conclusion of \cite{geha12} that no quenched galaxies with M$_* < $10$^9$M$_{\odot}$ are found in the field. Only among the nonisolated dwarfs (both paired and unpaired) do we find any quenched systems, thus indicating that the primary form of quenching in dwarfs is due to large scale environmental factors (for example, ram-pressure stripping) rather than interaction with other dwarfs.

\section{High Gas Fractions in Paired Dwarfs}\label{gasfrac}
Since quenched dwarf systems are not observed in isolation \citep{geha12}, yet SFRs are enhanced in isolated dwarf pairs (Figure \ref{sfrenhance}), the processes leading up to the coalescence stage in a galaxy merger are likely not as efficient at exhuasting the fuel for future star formation in dwarfs as they are in mergers between more massive galaxies. Differences in the final merger product may arise because of inefficient star formation in dwarfs at earlier stages of the interaction, or because the gas reservoirs for dwarf pairs may be distinct environments from those of isolated dwarfs without companions.

\begin{figure*}
\begin{center}
\includegraphics[height=2.3in,width=3.2in,viewport=20 0 500 355,clip]{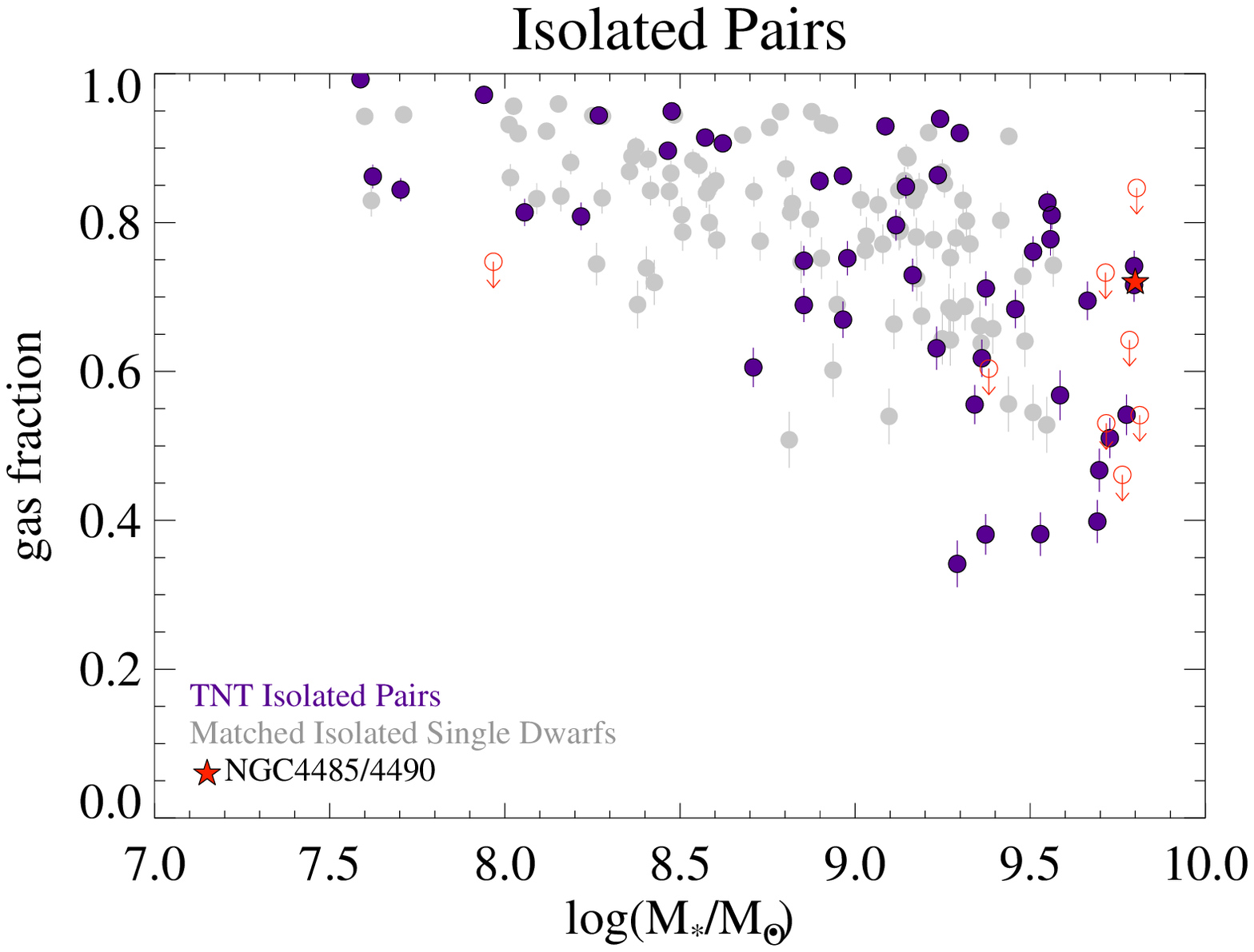}
\includegraphics[height=2.3in,width=3.2in,viewport=20 0 500 355,clip]{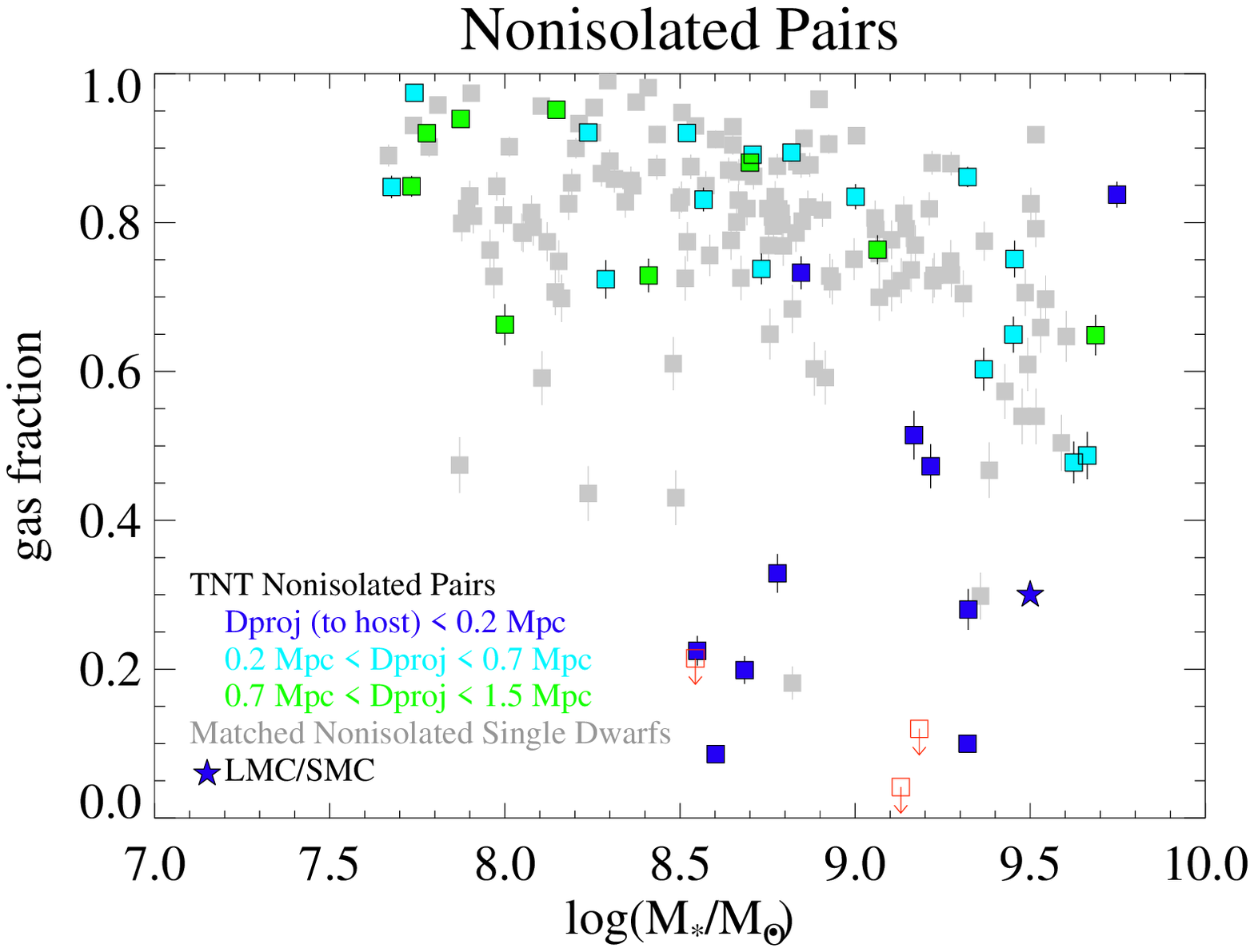}
\caption{Total atomic gas fraction ($f_{gas} = 1.4M_{HI}/(1.4M_{HI} + M_*)$) as a function of total stellar mass for TNT isolated (left) and nonisolated (right) dwarf pairs. In each panel, the gray points show the gas fractions for the matched isolated (left, gray circles) and nonisolated (right, gray squares) single dwarfs that are detected by ALFALFA. The colored points represent the gas fractions that probe the \HI\ gas in and around each pair for isolated dwarf pairs (left, purple circles) and for nonisolated dwarf pairs color coded by distance from a massive host (right, colored squares). Upper limits for the TNT pairs are shown in red. All samples have similarly high gas fractions (f$_{gas} > $0.6) except for nonisolated dwarf pairs within 200 kpc of a massive neighbor and a few isolated dwarf pairs at high stellar mass (M$_* > $10$^{9.2}$ M$_{\odot}$). Thus large scale environmental effects, rather than dwarf-dwarf interactions, appear to be the most effective mode of quenching in dwarfs. The interacting dwarf pairs NGC4485/4490 (left, red star) and the LMC/SMC (right, blue star) are marked for reference.
\label{gasfracenhance}}
\end{center}
\end{figure*}

We probe the neutral gas in and around the TNT pairs in Figure \ref{gasfracenhance}, where gas fractions, defined as $f_{gas} = 1.4M_{HI}/(1.4M_{HI} + M_*)$ (with a factor of 1.4 to account for Helium), are shown as a function of stellar mass for the isolated (left panel, purple circles) and nonisolated (right panel, colored squares) dwarf pairs. Since single-dish telescopes with large beam sizes were used to obtain the observations of the neutral gas (see Section \ref{HIreduct}), each colored point in Figure \ref{gasfracenhance} represents the total masses for the pair. Also shown in Figure \ref{gasfracenhance} are the gas fractions as a function of stellar mass for the matched isolated single dwarfs (left panel, gray circles) and for the matched nonisolated single dwarfs (right panel, gray squares). The stellar and \HI\ masses for the control sample come from the \cite{mendel} and \cite{a40} catalogs, respectively, as described in Section \ref{controls}. Only the matched controls that are detected in ALFALFA are plotted.

The left panel of Figure \ref{gasfracenhance} shows that the TNT isolated dwarf pairs follow a similar distribution of high gas fractions (f$_{gas} >$ 0.6) to their matched controls, despite having higher SFRs on average (as shown in Figure \ref{sfrenhance}). Only at stellar masses of log(M$_*$/M$_{\odot}$) $>$ 9.2 do some of the TNT isolated dwarf pairs have gas fractions lower than those of their matched isolated single dwarfs (f$_{gas} <$ 0.4). This stellar mass limit coincides with the mass at which quenched galaxies are observed in the field \citep{geha12} and thus may indicate a mass range within which dwarfs in isolation from massive neighbors can effectively lower their gas fractions by being in a dwarf-dwarf pair. The three pairs with log(M$_*$/M$_{\odot}$) $>$ 9.2 and f$_{gas} <$ 0.4 are at projected separations of 13-20 kpc. Two of the pairs with low f$_{gas}$ have high stellar mass ratios (M$_{*,1}$/M$_{*,2} >$ 4.5) and one pair member with an H$\alpha$ EQW $>$ 50\AA.

The importance of large scale environmental effects on dwarf galaxy evolution is evident in the right panel of Figure \ref{gasfracenhance}. Almost all pairs closer than 0.2 Mpc to a massive galaxy (dark blue squares) have observed gas fractions $<$0.6, the lower limit observed for most of the other nonisolated pairs. One exception is the outlier in the top right at high stellar mass and high gas fraction. Due to the large single-dish beam, this pair has an overestimated gas fraction that is contaminated by \HI\ from its massive neighbor. Among the dwarf pairs farther from their massive hosts, we still observe gas fractions that cover a similar range to that observed for the matched, nonisolated single dwarfs. Thus large scale environmental effects rather than dwarf-dwarf interactions appear to be the most effective mode of quenching in dwarfs.

There are eight TNT isolated pairs and three TNT nonisolated pairs for which we do not have firm \HI\ detections (red circles and squares in Figure \ref{gasfracenhance}) marked as upper limits by arrows). However, we do not expect any of these nondetections to affect the gas fraction distributions seen in Figure \ref{gasfracenhance}. As can be seen in Figure \ref{atlas}, four of the isolated pairs noted as nondetections are actually marginal detections at the 3-4$\sigma$ level and thus likely require only slightly longer integration times to measure reliable \HI\ parameters. Three of the four remaining nondetections for the isolated pairs are among the most distant in our sample with D $>$ 230 Mpc leading to \HI\ upper limits of M$_{HI} = 1-4\times$10$^{10}$M$_{\odot}$ (assuming a S/N ratio of 4 and a velocity width of 150 \kms) that are well above the median \HI\ mass we observe for our detections (M$_{HI} = $3$\times$10$^9$M$_{\odot}$). Thus, it is highly likely that, given slightly deeper integrations, these pairs would also be detected at similar \HI\ masses to those already sampled by the detections shown in Figure \ref{gasfracenhance}. Even if the \HI\ masses were ten times lower than these upper limits, the gas fractions would be reduced to f$_{gas}\sim$0.2-0.3, but all three of these nondetections are in the highest stellar mass bin where such low gas fractions are already represented. The three nondetections among the nonisolated pairs, however, have upper limits well within our sensitivities and place tight constraints on their gas fractions (f$_{gas}\sim$0.05-0.2). These three pairs are also within 100 kpc of a massive galaxy and thus are likely quenched systems.

One intriguing exception among the nondetections for the TNT isolated pairs is the nondetection with a total stellar mass for the pair of log(M$_*$/M$_{\odot}$) = 8.0 and total gas fraction for the pair of f$_{gas} < $0.75. This pair (shown third from the left in Figure \ref{atlas}) has the second smallest projected radial separation (R$_{sep} = $1.08 kpc) and a velocity separation of v$_{sep} = $42 /kms. If this system is a dwarf pair in the end stages of coalescence, the low gas fraction given its stellar mass could reflect the beginnings of quenching as a result of exhausting its gas supply via the merger. However, since we find only one such pair, with a low gas fraction relative to the matched controls and a low stellar mass (M$_* < $10$^{9.2}$), this process is not common in our sample. Alternatively, the two nuclei of the pair members that have begun to merge under a common stellar envelope may instead be two different HII regions within the same dwarf galaxy since the two morphological scenarios are difficult to discern at the SDSS resolution and without further kinematic information. 

\section{Discussion}\label{disc}
Approximately 10\% of all low mass galaxies are expected to have undergone a major merger since z$\sim$1 \citep{deason14}, compared to observed merger rates of only 3-6\% for their more massive counterparts \citep{patton6}. Given that dwarfs also outnumber massive galaxies in the local universe \citep{tullycat, karacat} and at higher redshift \citep{fakhouri}, interactions between dwarf galaxies likely present an important mode of star formation and a significant role in processing baryons in the universe. Under the framework of hierarchical structure formation, this pre-processing in dwarfs (i.e. processing of their gas before they are accreted by a massive host) has significant implications for the assembly history of galaxies over cosmic time. 

\subsection{Placing Local Dwarf Pairs into Context with TNT}
TNT is the first systematic study of a large sample of dwarf pairs caught in various stages of interaction, which provides the first opportunity to place the small number of individual examples of known dwarf-dwarf interactions into the larger context of the dwarf-dwarf merger sequence. As discussed in Section \ref{knownpairs}, the primary member in the NGC4485/4490 pair, NGC4490, is slightly more massive (M$_* \sim$ 10$^{10}$M$_{\odot}$) than the TNT selection criterion. However, the pair still offers a valuable comparison as a rare example of a very nearby (D$=$ 8 Mpc), clearly interacting pair of low mass galaxies found in isolation from a massive neighbor. As shown in Figure \ref{sbproperties}, starbursts are induced over the full range of radial separations probed by the main TNT sample and most often in the secondary (less massive) pair member, but only at velocity separations of v$_{sep} <$ 70 \kms. At a projected distance from NGC4490 of R$_{sep} \sim $10 kpc, NGC4485, which is undergoing a global starburst \citep{11HUGS}, is well within the range of projected radial separations for which triggered starbursts are observed in TNT. However at v$_{sep} =$ 72 \kms, the pair represents the upper limit to the velocity separations observed for pairs containing at least one starbursting member. As shown in Figure \ref{gasfracenhance}, the atomic gas fraction for the NGC4485/4490 pair, which is known to be embedded in a large, diffuse \HI\ component \citep{clemens98}, is similar at f$_{gas} =$ 0.72 to the fractions observed for other TNT isolated pairs. Thus, such large \HI\ envelopes may be common for dwarf pairs in isolation.

Analogs to the only example of an interacting dwarf pair in the Local Group, the Large and Small Magellanic Clouds, are well-represented in the TNT nonisolated pair sample in terms of R$_{sep}$, v$_{sep}$, and M$_*$. Despite the fact that the LMC is bluer on average than its analogs in the field \citep{tollerudLMC}, neither the LMC nor the SMC are starbursts. The lack of triggered starbursts in the pair is consistent with the approximate limits in projected radial and velocity separation revealed in Figure \ref{sbproperties} for the TNT nonisolated pairs. At R$_{sep} \sim $ 15 kpc, the LMC/SMC pair sits at the largest pair separation observed for TNT nonisolated pairs containing at least one starbursting member. The line of sight velocity separation of the pair, v$_{sep} = $ 120 \kms, however, is well above the upper limit of v$_{sep} \sim$ 70 \kms\ found for starbursting TNT nonisolated pairs. The LMC/SMC have a low atomic gas fraction (f$_{gas} =$ 0.3) consistent with other pairs with similar proximity to a massive host (D $<$ 200 kpc). We discuss possible reasons for these low gas fractions in the next section. 

Finding dwarf pairs with high mass ratios is challenging in SDSS due to the sensivitity of the spectroscopic survey to dwarf companions, a strong function of redshift. However, it is not impossible, as evidenced by the three isolated pairs and four nonisolated pairs in TNT with M$_{*,1}$/M$_{*,2} > $ 7.5. Despite the fact that both the NGC4485/4490 and LMC/SMC pairs have M$_{*,1}$/M$_{*,2} > $ 7.5, they are not otherwise outliers in Figures \ref{sbproperties} or \ref{gasfracenhance}. Thus, at least in terms of their starburst properties and atomic gas fractions, high mass ratio pairs like NGC4485/4490 and LMC/SMC show results consistent with those of the lower mass ratio (i.e. major) mergers that make up most of the TNT sample.

\subsection{The Impact of Dwarf-Dwarf Interactions on the Baryon Cycle}
Using the metrics of star formation, triggered starbursts, and gas fractions, TNT reveals the dwarf-dwarf merger sequence has pronounced consequences on the evolution of low mass galaxies. Star formation is enhanced by a factor of 2.3 ($\pm$0.7) in dwarf pairs with R$_{sep} <$ 50 kpc. The lack of similarly high SFRs in the unpaired systems suggests the additional star formation is due to interactions and not stochastic effects. Prior encounters between pair members could help explain SFR enhancements observed in pairs as far apart as roughly a virial radius (R$_{vir} \sim$ 100 kpc for a galaxy of M$_* \sim$ 10$^9$M$_{\odot}$). 
Interaction-driven star formation is not expected to shut off immediately after a first close passage between two massive galaxies in a merging system, but instead should continue to increase for $\sim$0.5 Gyr as the pair members move apart from one another and then remain enhanced for even longer \citep{torrey12,patton13}. Similar detailed modeling has not been done explicitly for mergers of galaxies with shallower gravitational potentials. However, given a sufficiently high orbital eccentricity, star formation initiated by a close encounter could plausibly continue over the timescales needed to reach such large pair separations.

Starbursts triggered by massive galaxy interactions are often observed in the later stages of coalescence when the two galaxies begin to merge under one common stellar envelope. Within this context, the population of starbursting, blue compact dwarf galaxies (BCDs), as well as other starbursting but seemingly isolated dwarfs, have been noted as the potential end stages of dwarf-dwarf interactions \citep[e.g.,][]{lelliSB1, lelliSB2, kolevaSB, IC10SB}. However, by sampling the full dwarf-dwarf merger sequence, TNT has shown that starbursts can occur across the full range of projected separations probed by our main sample (R$_{sep} <$ 50 kpc) and thus at much earlier stages in the interaction. These bursts do, however, prefer pairs with velocity separations $<$70 \kms. The escape speed at 50 kpc for a dwarf with M$_*\sim$10$^9$M$_{\odot}$ (and thus M$_{halo}\sim$10$^{11}$M$_{\odot}$) is $\sim$130 \kms. Thus these low velocity separations suggest the starbursting pairs are not undergoing fly-by interactions but instead, given similarly low transverse velocities, are bound to one another, and a prior encounter could have triggered the starbursts in the more widely separated pairs.

\begin{figure*}
\begin{center}
\includegraphics[height=2.3in,width=3.2in,viewport=20 0 500 355,clip]{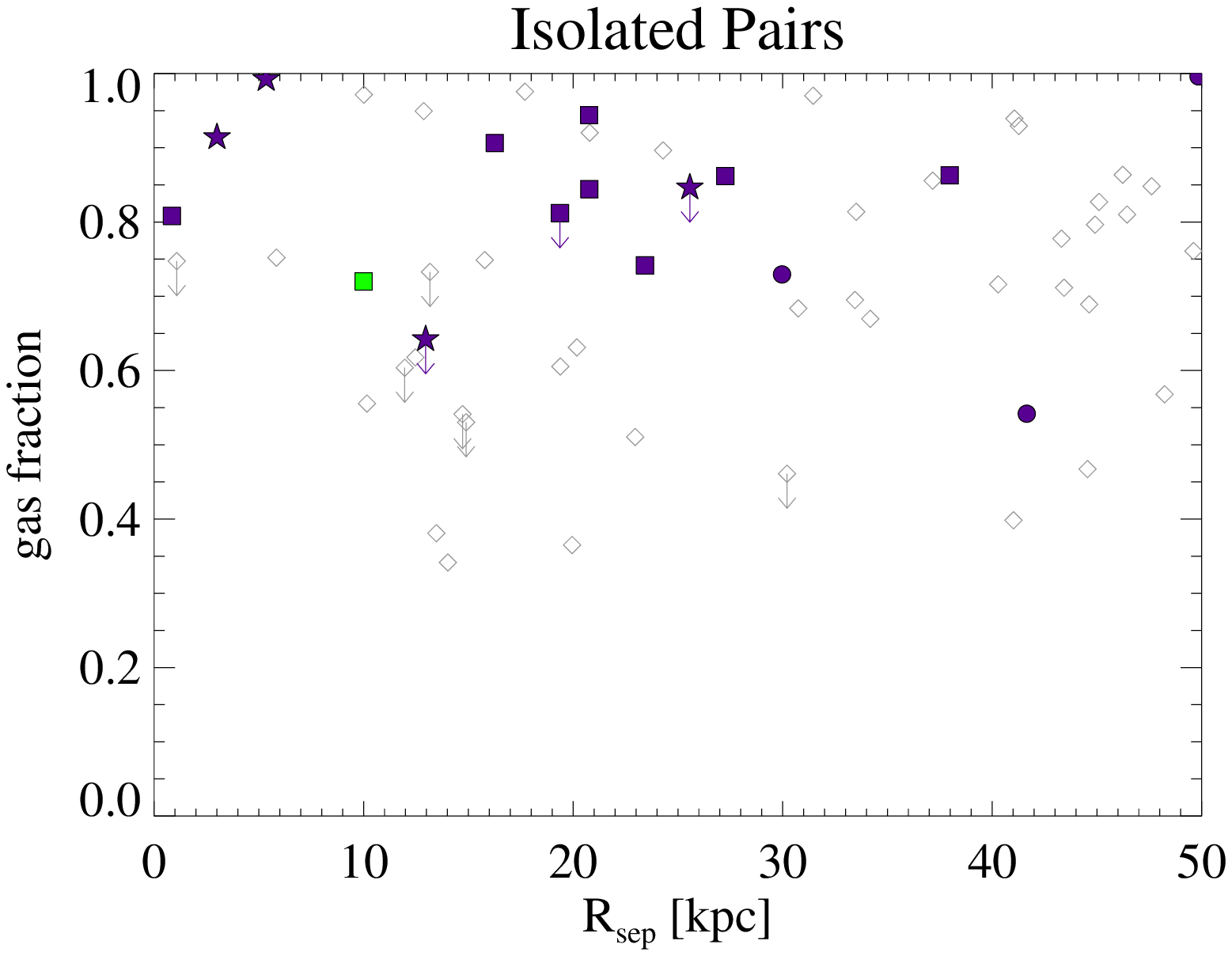}
\includegraphics[height=2.3in,width=3.2in,viewport=20 0 500 355,clip]{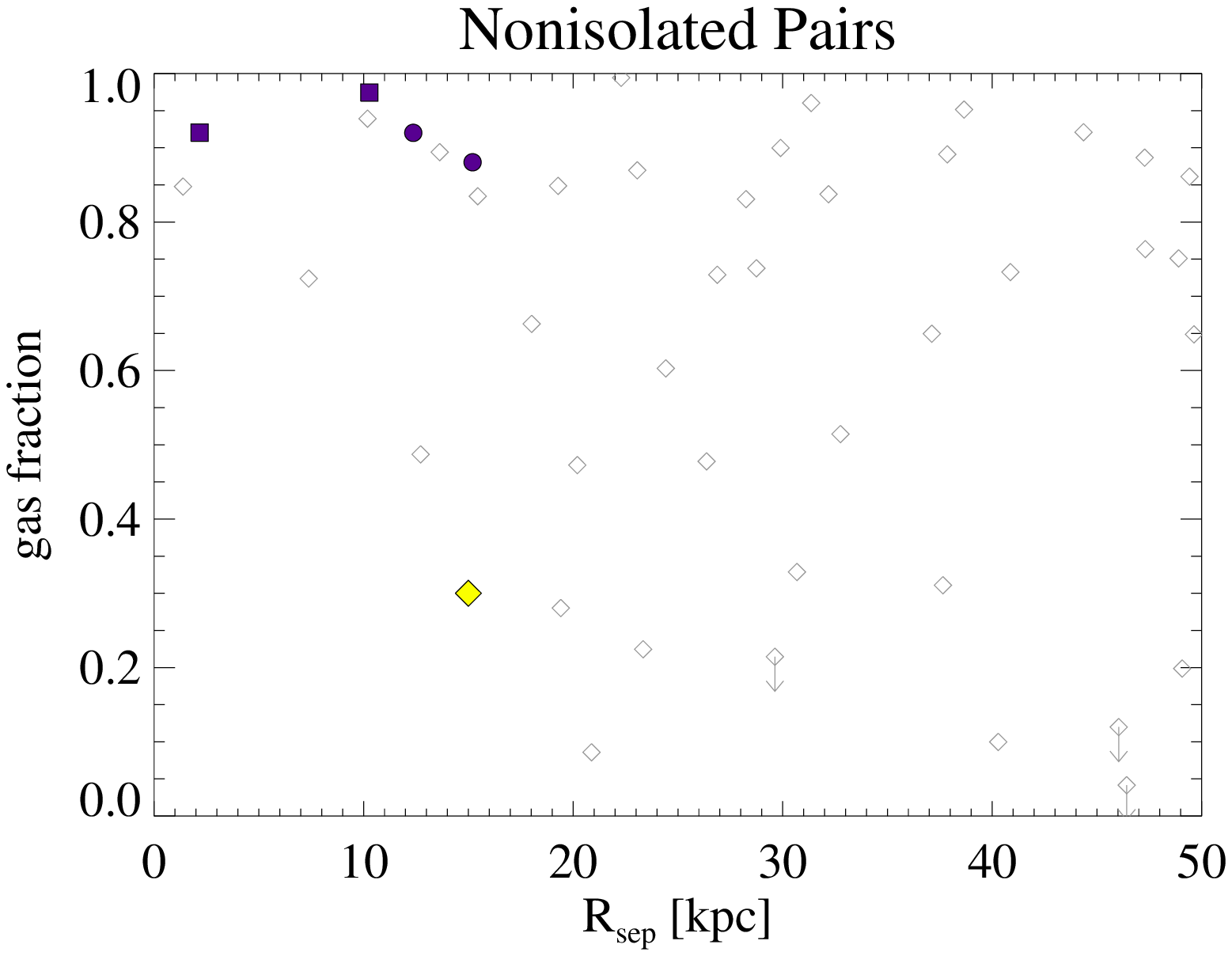}
\caption{Atomic gas fractions (f$_{gas} =$ 1.4M$_{HI}$/(1.4M$_{HI} +$ M$_*$)) that probe gas in and around each dwarf pair as a function of projected radial separation between pair members for TNT isolated (left panel) and nonisolated (right panel) dwarf pairs. As in Figure \ref{sbproperties}, we mark for each pair whether the primary (circle), secondary (square), or both (star) pair members are starbursting, and nonstarburst pairs are shown as gray diamonds. A dwarf pair member is considered starbursting if its SFR is more than five times the average SFR for dwarfs of similar stellar mass, using the star formation main sequence (i.e. stellar mass vs. average SFR relation) of \cite{luo} (see Column 3 of Table 1). Atomic gas fraction does not vary with pair separation, and pairs hosting starbursts are on average more gas rich. The well-studied interacting pair NGC4485/NGC4490 (green square) and the LMC/SMC (yellow diamond) are marked for reference.
\label{rsepfgas}}
\end{center}
\end{figure*}

The strong SFR enhancements and triggered starbursts among both the TNT isolated and nonisolated dwarf pairs strongly suggest that tidally triggered star formation plays an important role in the evolution of these low mass galaxies. The high SFRs would seem to be consistent with lower than average gas masses, in keeping with expectations from more massive counterparts that eventually experience quenching of their gas supplies. However, almost all of our pairs, both isolated and nonisolated, have large gas reservoirs (M$_{HI} > 2\times$10$^8$ M$_{\odot}$) and gas fractions similar to those observed for matched, single galaxies of similar stellar mass (Figure \ref{gasfracenhance}). In fact, the dwarf pairs containing at least one starbursting member also typically have the highest gas fractions. As shown in Figure \ref{rsepfgas}, starbursts are only triggered in the TNT isolated pairs with f$_{gas} \gtrsim$ 0.6, and the cutoff is even higher among the TNT nonisolated pairs: only pairs with f$_{gas} \gtrsim$ 0.85 contain starbursts.

These untapped gas reservoirs may indicate that dwarf pairs preferentially form in very gas-rich environments, and thus their gas content remains high even if some of that gas is lost to the interaction, either through tidal stripping or feedback, or as fuel for starbursts. If diffuse gas surrounding the galaxies dominates the neutral gas content (i.e. tidal stripping is efficient), then, even if star formation is augmented in the main galaxy, there may still be a large amount of neutral gas sitting outside of the interacting galaxies and not forming stars. The additional \HI\ in these environments may, in turn, more easily shield the \HI\ gas from ionizing UV background, thus allowing more gas to remain in the neutral phase. This diffuse component may later fall back into the galaxies, triggering future star formation and thus further delaying quenching. Merger simulations of Milky Way-like systems \citep{hopkins13} and even galaxies down to SMC-type masses \citep{renaud13, renaud14} show quenching to be gradual; feedback from star formation pushes gas out of the galaxy only to have it fall back in and thus slow the cessation of star formation. The same effect may be at work to maintain the high gas fractions observed for the dwarf pairs which have average gas depletion times of t$_{dep} =$ M$_{HI}$/SFR$_{tot} \sim$ 10 Gyr.  

Both the LMC/SMC and the NGC4485/4490 pairs have a significant neutral gas component outside of the galaxies themselves \citep{putmanstream,clemens98}. A large VLA program ($>$180 hours) to map the \HI\ emission in a subset of the TNT dwarf pairs is currently underway and preliminary results show between 30-50\% of the \HI\ component exists in an extended, diffuse component (Stierwalt et al. $in~prep$). Cosmological hydrodynamics simulations also suggest that mergers are found in more \HI\ gas-rich environments than their nonmerging counterparts \citep{rafi}.

Only at the highest (total) stellar masses (M$_* >$ 2$\times$10$^9$ M$_{\odot}$) for the isolated pairs do we observe any pairs with low gas fractions given their stellar mass. As discussed in Section \ref{gasfrac}, this stellar mass limit is also the minimum M$_*$ for which quenched galaxies are observed in the field \citep{geha12}, and thus may indicate a mass range within which dwarfs in isolation from massive neighbors can effectively lower their gas fractions by being in a dwarf-dwarf pair. However, not all the pairs with M$_* >$ 2$\times$ 10$^9$M$_{\odot}$ have lower gas fractions (Figure \ref{gasfracenhance}), and there is evidence that gas fractions remain constant over a range of projected pair separations for galaxy pairs at higher stellar masses (Fertig et al. $submitted$). 

Other, less likely, explanations for the combination of high SFRs and unquenched gas fractions could be short-lived bursts of star formation or underestimates in stellar mass. If star formation in paired dwarfs occurs in strong but short bursts, the gas reservoirs would remain mostly untapped. However, the existence of a SFR enhancement out to pair separations as wide as 100 kpc (see Figure \ref{sfrenhance}) contradicts the possibility of short bursts. Additionaly, high resolution studies of starbursts in nearby dwarfs have found burst durations as long as 450-650 Myr \citep{SBmcquinnI,SBmcquinnII}.

Our gas fractions are sensitive to large amounts of neutral gas outside of the galaxies themselves, but not to diffuse stellar material that may have been similarly removed, since we derive \HI\ mass from a measure of the global neutral gas content for the pair as a whole (due to large single-dish radio beam sizes). However, as shown in Figure \ref{gasfracenhance}, the \HI\ masses are in many cases 10$\times$ higher than the stellar masses and thus dominate the gas fractions. In order to lower a gas fraction from 0.9 to 0.7, for example, the stellar mass would have to be underestimated by a factor of four, leaving 3$\times$ the mass of the galaxies themselves in external tidal debris, a highly unlikely scenario. From simulations of the LMC/SMC, the stellar debris associated with the system is expected to be $\sim$7$\times$10$^7$ M$_{\odot}$ or on the order of up to one third the mass of the SMC \citep{besla13}.

\subsection{Dwarf-Dwarf Interactions in the Large Scale Environment}

Large scale environmental effects play a significant role in the baryon cycle of dwarf-dwarf interactions. When in proximity to a massive neighbor, interacting dwarfs have to be much closer together than their isolated counterparts to show enhanced star formation ($<$ 50 kpc; Figure \ref{sfrenhance}) or to trigger starbursts ($<$ 15 kpc; Figure \ref{sbproperties}). Many of our dwarf pair members show star-forming knots near their outer edges, much like those observed at high resolution in NGC4485 \citep{4485HST}. This mode of star formation may be prevented when environmental quenching is efficient. At larger pair separations, the fuel for star formation may be ram-pressure stripped by the massive neighbor. Similarly, the hot haloes that surround massive galaxies like the Milky Way could ionize the diffuse gas surrounding nonisolated dwarf pairs and thus make it inaccessible to star formation. Much of the gas surrounding the LMC/SMC is in an ionized state \citep{fox14}, thus possibly explaining the low atomic gas fraction observed for the pair (f$_{gas} =$ 0.3; Figure \ref{gasfracenhance}). If the mass of the ionized gas attributed to the Clouds is included in the gas fraction, the total gas mass becomes M$_{gas} \sim$ 2$\times$10$^9$ M$_{\odot}$ \citep[assuming a distance of 55 pc;][]{fox14} and the gas fraction would rise to 0.57. Using instead the distance of 100 kpc to the Magellanic Stream \citep{besla10}, the gas fraction becomes average (f$_{gas} \sim$ 0.7) compared to the other high gas fractions in Figure \ref{gasfracenhance}. 

Although there is a suggestion that gas depletion might be happening in some of the pairs at the highest stellar masses probed by TNT, the only clearly effective mechanism for quenching any of the dwarfs in our study is proximity to a massive neighbor. Only the pairs closest to a massive galaxy (D$<$ 200 kpc) show low gas fractions given their stellar mass and only for the nonisolated systems do we find any dwarfs (paired or unpaired) with H$\alpha$ EQW $<$ 2\AA\ (Table 1).

\section{Conclusions}
We introduced TiNy Titans (TNT), the first systematic study of star formation and the subsequent processing of the ISM in interacting dwarfs as a population. Through both a multiwavelength observational effort to study a complete sample of interacting dwarfs and a theoretical approach using cosmological simulations and hydrodynamic simulations of individual systems, TNT will explore aspects of dwarf-dwarf interactions as a mode of dwarf galaxy evolution not possible with a small number of local examples. Specifically, we aim 1) to establish the dwarf-dwarf merger sequence at z$=$0; 2) to connect interacting dwarfs at the current epoch with high redshift analogs; and 3) to probe interaction-driven modes of star formation.

The work presented here reflects the observational results toward the first goal of TNT: establishing the dwarf-dwarf merger sequence at z$=$0. We showed that dwarf-dwarf interactions occur both in isolation and near massive galaxy hosts and that these interactions have pronounced consequences for the star formation of the dwarfs involved. Specifically we found:
\begin{enumerate}
\item{Given the dwarf galaxy selection criteria of 10$^7$ M$_{\odot} <$ M$_* < 5 \times$10$^9$ M$_{\odot}$ and 0.005 $<$ z $<$ 0.07 and the pair selection criteria of R$_{sep} < 50$ kpc, v$_{sep} < $300 \kms, $(M_1/M_2)_* < $10, and D $>$ 1.5 Mpc from a massive neighbor, 60 isolated dwarf pairs are selected from the spectroscopic portion of the SDSS DR7. The pairs span the full range of R$_{sep}$, but most ($>$90\%) cover only v$_{sep} <$ 150 \kms\ and $(M_1/M_2)_* < $ 5.}
\item{Star formation rates in paired dwarfs are enhanced by a factor of 2.3 ($\pm$ 0.7) over isolated single dwarfs matched in redshift, stellar mass and local density for pair separations $<$ 50 kpc. The enhancement, which decreases with increasing pair separation, is observed out to pair separations as far as 100 kpc for isolated pairs. A similar enhancement is observed in the SFRs of more massive galaxy pairs, but the massive galaxy enhancement is not as large as that observed for dwarfs and extends to comparable pair separations despite the much larger extents of massive galaxies. The lack of similarly high SFRs in unpaired dwarfs suggests the enhancement is interaction-driven and not due to stochastic effects. Merger simulations of more massive galaxies suggest that prior close encounters trigger enhanced star formation which continues as the pair separates to such large distances.} 
\item{Starbursts, defined by H$\alpha$ EQW $>$ 100\AA, occur in 20\% of the TNT isolated pairs and in 20\% of the TNT nonisolated pairs compared to only 6\% (8\%) of the matched isolated (nonisolated) single dwarfs. Thus, dwarf-dwarf interactions are an effective mode of producing starbursts in dwarfs. Our starburst statistics for the unpaired dwarfs match those of \cite{11HUGS} who performed a census of contributions from dwarf starbursts in the local volume.}
\item{Among the TNT isolated pairs, starbursts are observed out to projected pair separations of 50 kpc showing that starbursts can be triggered throughout a merger, including early stages in the interaction, and not just near coalescence. Starbursts are only hosted in isolated pairs with v$_{sep} <$ 70 \kms\ which is much lower than the escape velocity of $\sim$130 \kms\ indicating they are not fly-by interactions but instead could be bound systems. Starbursts are more commonly triggered in the secondary (i.e. lower mass) member of the pair, and most pairs for which the primary or both the primary and secondary are starbursting have mass ratios of 1-2. }
\item{Despite their enhanced SFRs and triggered starbursts, the TNT isolated dwarf pairs have similar gas fractions relative to isolated single dwarfs at the same stellar mass. Only a few of the isolated pairs with log(M$_*$/M$_{\odot}$) $>$ 9.2, i.e. the same minimum mass above which quenched galaxies are observed in the field \citep{geha12}, show signs of possible gas depletion. Thus, there may be a diffuse component contributing to the neutral gas content that sits outside of the galaxies themselves and is not forming stars like the one observed for the interacting pair NGC4485/4490 \citep{clemens98}. Star formation also may not have been enhanced for long enough to significantly deplete the gas.}
\item{We observe no trends in atomic gas fraction with projected pair separation for the either the TNT isolated or nonisolated pairs. However, in both samples the pairs containing at least one starbursting member all have high gas fractions (f$_{gas} >$ 0.6 for isolated pairs and f$_{gas} >$ 0.85 for nonisolated pairs). Thus systems that are $very$ gas-rich, even relative to the already gas-rich dwarfs in our sample, are required to trigger starbursts which are then inefficient at burning through that fuel.}
\item{Proximity to a massive galaxy neighbor has clear effects on dwarf-dwarf interactions. Star formation is enhanced by a factor of 2.3 ($\pm$ 0.4) in the TNT nonisolated pairs relative to their unpaired (and also nonisolated) counterparts, but only to pair separations R$_{sep} <$ 50 kpc. Starbursts are triggered in nonisolated dwarf pairs, but not at pair separations $>$ 15 kpc (the pair separation of the LMC/SMC). Thus, paired dwarfs must be much closer to one another, i.e. more progressed in their interaction, for the effects of the interaction to dominate over those of the larger scale environment. Dwarfs, both paired and unpaired, with H$\alpha$ EQW $<$ 2 \AA\ are only found within 1.5 Mpc of a more massive galaxy, and the only dwarf pairs with f$_{gas} <$ 0.4 are within 200 kpc of a massive host. Thus, proximity to a massive host, and not the processing of gas via star formation enhanced by a dwarf-dwarf interaction, is the only effective mode for quenching in dwarfs.}
\item{The star formation histories and gas fractions observed for the LMC/SMC and NGC4485/4490 interacting dwarf pairs are consistent within the context of the merger sequence described by TNT, including the large gas reservoir surrounding the NGC4485/4490 pair and the fact that neither the LMC nor SMC are starbursting despite enhanced SFRs.}
\end{enumerate}

Higher resolution imaging of the gas is necessary to understand the dominant processes at work in interacting dwarfs to maintain both their high SFRs and high gas fractions. We currently have a large program using the Karl G. Jansky Very Large Array (PI: Stierwalt) to map the neutral hydrogen in a subset of the TNT pairs to determine whether tidal stripping has occurred in these interactions and what fraction of the gas exists in a neutral component, outside of the galaxies. To asses total gas content, we have an ongoing program to obtain single dish CO measurements for each pair member (PI: Stierwalt). An ongoing deeper optical imaging campaign will also reveal whether stellar material is also being stripped, as well as reveal where most of the triggered star formation is taking place within each dwarf (PIs Johnson, Kallivayalil, \& Stierwalt). These multiwavelength resolved studies will also aid in future theoretical pursuits, specifically star formation prescriptions and the orbital properties of dwarf groups.\\

The TNT Team thanks the entire ALFALFA collaboration whose many hours of observing and data reduction provided us the \HI\ measurements for two of our control samples which were critical to this work. The authors also thank M. Geha and J. Bradford for many helpful and productive discussions leading up to and throughout the writing of this paper, as well as D.J. Pisano \& T. Osterloo for their advice on GBT calibration. S. Stierwalt, G. Besla, \& N. Kallivayalil thank the wonderful people at the Aspen Center for Physics and the NSF Grant \#1066293 for their hospitality during the writing of this paper. S. Stierwalt thanks J. Lee, A. Leroy, A. Evans, and S. Ellison for their thoughts and advice. S. Stierwalt and K.E.Johnson thank the David and Lucile Packard Foundation for generous support through a Packard Fellowship. S. Stierwalt also gratefully acknowledges the L'Oreal USA For Women in Science program for their grant to conduct this research.

This work has made use of the NASA/IPAC Extragalactic Database (NED) which is operated by the Jet Propulsion Laboratory, California Institute of Technology, under contract with the National Aeronautics and Space Administration. 

This work has also used catalogs and imaging from the NASA Sloan Atlas and the SDSS. Funding for the NASA-Sloan Atlas has been provided by the NASA Astrophysics Data Analysis Program (08-ADP08-0072) and the NSF (AST-1211644). Funding for the SDSS and SDSS-II has been provided by the Alfred P. Sloan Foundation, the Participating Institutions, the National Science Foundation, the U.S. Department of Energy, the National Aeronautics and Space Administration, the Japanese Monbukagakusho, the Max Planck Society, and the Higher Education Funding Council for England. The SDSS Web Site is http://www.sdss.org/. The SDSS is managed by the Astrophysical Research Consortium for the Participating Institutions. The Participating Institutions are the American Museum of Natural History, Astrophysical Institute Potsdam, University of Basel, University of Cambridge, Case Western Reserve University, University of Chicago, Drexel University, Fermilab, the Institute for Advanced Study, the Japan Participation Group, Johns Hopkins University, the Joint Institute for Nuclear Astrophysics, the Kavli Institute for Particle Astrophysics and Cosmology, the Korean Scientist Group, the Chinese Academy of Sciences (LAMOST), Los Alamos National Laboratory, the Max-Planck-Institute for Astronomy (MPIA), the Max-Planck-Institute for Astrophysics (MPA), New Mexico State University, Ohio State University, University of Pittsburgh, University of Portsmouth, Princeton University, the United States Naval Observatory, and the University of Washington.
 
\bibliography{ddmreferences}{}
\end{document}